\documentclass[twocolumn]{emulateapj}  
\usepackage[left]{lineno}
\usepackage{epstopdf}
\usepackage{ulem}
\usepackage{amssymb}
\usepackage{times}
\usepackage{graphicx}
\usepackage[usenames,dvipsnames]{pstricks}
\usepackage{psfrag}
\usepackage{lipsum}
\usepackage{natbib}
\usepackage[colorlinks=true,linkcolor=blue,citecolor=blue]{hyperref}
\def\araa{ARA\&A}
\def\apj{ApJ}
\def\apjl{ApJ}
\def\apjs{ApJS}
\def\aap{A\&A}
\def\aaps{A\&AS}
\def\mnras{MNRAS}
\def\nat{Nature}
\def\physrep{Phys.~Rep.}

\newcommand{\be}{\begin{equation}}
\newcommand{\ee}{\end{equation}}
\newcommand{\bary}{\begin{eqnarray}}
\newcommand{\eary}{\end{eqnarray}}

\begin{document}

\title{Signatures from a quasi-spherical outflow and an off-axis top-hat jet launched in a\\ merger of compact objects: An analytical approach}

\author{ N.~ Fraija\altaffilmark{1$\dagger$}, D.~ Lopez-Camara\altaffilmark{2}, A.C. Caligula do E. S. Pedreira\altaffilmark{1},  B.~ Betancourt Kamenetskaia  \altaffilmark{1}, P. ~ Veres\altaffilmark{3} and S.~ Dichiara$^{4,5}$ 
}
\affil{$^1$Instituto de Astronom\'ia, Universidad Nacional Aut\'{o}noma de M\'{e}xico, Apdo. Postal 70-264, Cd. Universitaria, Ciudad de M\'{e}xico 04510
}
\affil{$^2$ CONACyT - Instituto de Astronom\'ia, Universidad Nacional Aut\'{o}noma de M\'{e}xico, Apdo. Postal 70-264, Cd. Universitaria, Ciudad de M\'{e}xico 04510
}
\affil{$^3$ Center for Space Plasma and Aeronomic Research (CSPAR), University of Alabama in Huntsville, Huntsville, AL 35899, USA
}
\affil{$^4$ Department of Astronomy, University of Maryland, College Park, MD 20742-4111, USA\\
$^5$ Astrophysics Science Division, NASA Goddard Space Flight Center, 8800 Greenbelt Rd, Greenbelt, MD 20771, USA
}
\email[$\dagger$ ]{nifraija@astro.unam.mx}

\begin{abstract}
The production of both gravitational waves and short gamma-ray bursts (sGRBs) is widely associated with the merger of compact objects. Several studies have modelled the evolution of the electromagnetic emission using the synchrotron emission produced by the deceleration of both a relativistic top-hat jet seen off-axis, and a wide-angle quasi-spherical outflow (both using numerical studies). In this study we present an analytical model of the synchrotron and synchrotron self-Compton (SSC) emission for an off-axis top-hat jet and a quasi-spherical outflow. We calculate the light curves obtained from an analytic model in which the synchrotron and SSC emission (in the fast- or slow-cooling regime) of an off-axis top-hat jet and a quasi-spherical outflow are decelerated in either a homogeneous or a wind-like circumburst medium. We show that the synchrotron emission of the quasi-spherical outflow is stronger than that of the off-axis jet during the first $\sim$ 10 - 20 days, and weaker during the next $\gtrsim$ 80 days. Moreover, we show that if the off-axis jet is decelerated in a wind-like medium, then the SSC emission is very likely to be detected. Applying a MCMC code to our model (for synchrotron emission only), we find the best-fit values for the radio, optical and X-ray emission of GRB 170817A which are in accordance with observations.  For GRB 170817A,  we find using our model that the synchrotron emission generated by the quasi-spherical outflow and off-axis top-hat jet increase as $F_\nu\propto t^{\alpha}$ with $\alpha\lesssim 0.8$ and $\alpha>3$, respectively.  Finally, we obtain the correspondent SSC light curves which are in accordance with the very-high-energy gamma-ray upper limits derived with the GeV - TeV observatories. 
\end{abstract}

\keywords{Gamma-rays bursts: individual (GRB 170817A) --- Stars: neutron --- Gravitational waves --- Physical data and processes: acceleration of particles  --- Physical data and processes: radiation mechanism: nonthermal --- ISM: general - magnetic fields}

\section{Introduction}

The merger of two neutron stars (NSs) is believed to be a potential candidate for the production of both gravitational waves (GWs), and short gamma-ray bursts (sGRBs, $T_{90}\lesssim 2$ s) with an extended non-thermal emission \citep[for reviews, see][]{2007PhR...442..166N, 2014ARA&A..52...43B}.  Since the first detection of a sGRB, GWs had been exhaustively searched without success.

On August 17, 2017 for the first time, a GW source \citep[GW170817;][]{PhysRevLett.119.161101,2041-8205-848-2-L12} was associated with a faint electromagnetic $\gamma$-ray counterpart which is most probably the prompt emission of a sGRB \citep[GRB170817A;][]{2017ApJ...848L..14G, 2017ApJ...848L..15S},  although today an off-axis emission from the top-hat jet is not the common knowledge \citep[e.g., see][]{2019MNRAS.486.1563M}.

GRB 170817A was followed up by an enormous observational campaign covering a  large fraction of the electromagnetic spectrum.  The X-ray counterpart was detected by the Chandra and XMM-Newton satellites \citep{troja2017a, 2018arXiv180103531M, 2018arXiv180502870A, 2018arXiv180106164D, 2017ATel11037....1M, 2018ATel11242....1H}. In optical bands, the non-thermal optical afterglow emission was revealed by the Hubble Space Telescope \citep{2018arXiv180102669L, 2018arXiv180103531M}.\footnote{By optical we refer to the optical afterglow emission rather than the kilonova emission.} The radio emission at 3 and 6 GHz was identified by the Very Large Array \citep[][]{2041-8205-848-2-L12, 2017arXiv171111573M, 2018ApJ...858L..15D, 2017Natur.547..425T}. The GW170917/GRB 170817A event was also within the field of view of the Large Area Telescope (LAT) on-board the Fermi satellite and the field of view of two of the TeV $\gamma$-ray observatories: the High Energy Stereoscopic System  (H.E.S.S.) Telescope and the High Altitude Water Cherenkov (HAWC). Fermi-LAT began observing around the position of GW170817 at $\sim$ 1000 s after the GBM trigger \citep{2041-8205-848-2-L12}. No counts were registered and upper limits were derived. Observations with the HAWC observatory began on 2017 August 17 at 20:53 UTC and finished 2.03 hr later \citep{2017GCN.21683....1M}. Although no significant excess was detected, upper limits for energies larger than 40 TeV were placed. The H.E.S.S. telescope searched for very-high-energy $\gamma$-ray emission in two opportunities \citep{2017ApJ...850L..22A}. First, 5.3 h after the GBM trigger, and secondly from 0.22 to 5.2 days after the trigger. No statistically significant excess of counts were found by this TeV observatory and upper limits were derived.\\

The temporal behaviour of the electromagnetic (EM) counterpart of GW170817 was atypical. The extended X-ray and radio afterglow were initially described by a simple power law which gradually increased as $\sim t^{0.8}$ \citep{2017arXiv171111573M, 2018arXiv180103531M}, reached its peak at $\sim$140 days after the NS merger, and was then followed by relatively fast  decline. A miscellaneous set of models based on external shocks such as off-axis top-hat jets \citep{troja2017a,2017ApJ...848L..20M, 2017arXiv171005905I, 2017ApJ...848L..21A, 2019ApJ...871..123F,2019arXiv190600502F}, radially stratified ejecta \citep{2017arXiv171111573M, 2019ApJ...871..200F,2018ApJ...867...95H} and structured jets \citep{2017Sci...358.1559K, 2017MNRAS.472.4953L, 2017arXiv171203237L} were developed to interpret the behavior of this atypical afterglow. \cite{2018Natur.561..355M} reported the detection of superluminal apparent motion in the radio band using Very Long Baseline Interferometry (VLBI), thus favoring models with successful jets and their respective quasi-spherical outflows. The successful jet models were also favored by VLBI radio observations performed by \citet{2018arXiv180800469G} at 207.4 days. In the latter, the authors constrained the size of the source, indicating that GRB 170817A was produced by a structured relativistic jet. The structured jet models \citep{2017Sci...358.1559K, 2017MNRAS.472.4953L, 2017arXiv171203237L}  suggest that the early-time radio flux during the two weeks after the merger is mostly produced by the radiation of an optically thin quasi-spherical outflow, while the late radio flux is dominated by the emission of a relativistic and collimated jet (immersed in the quasi-spherical outflow which is now optically thin) with an opening angle less than $\lesssim 5^\circ$ and observed from a viewing angle of $20^\circ\pm5^\circ$.\\

Fermi-LAT has detected  more than 50 GRBs with photons above 100 MeV and $\sim$12 bursts with photons above 10 GeV \citep[see][and references therein]{2013ApJS..209...11A,2014Sci...343...42A}. Several authors have suggested that although  photons with energies larger than $\geq 100$ MeV can be explained in the framework of the synchrotron forward-shock model  \citep{2009MNRAS.400L..75K, 2013ApJ...771L..33W, 2016ApJ...818..190F,2019arXiv190513572F,2019ApJ...879L..26F, 2019arXiv190706675F},  the maximum photon energy in this process is  $\sim 10~{\rm GeV}~\left(\frac{\Gamma}{100}\right)\left(1+z\right)^{-1}$, where $\Gamma$ is the bulk Lorentz factor and $z$ the redshift. Furthermore, taking into account that the bulk Lorentz factor during the afterglow phase evolves as $\propto t^{-\frac38}$ and $\propto t^{-\frac14}$ for a homogeneous and a wind-like medium, respectively,  GeV photons from synchrotron radiation are not expected at timescales of $\sim$ 100 s. Recently, the MAGIC collaboration reported the detection of 300-GeV photons for almost 20 minutes in the direction of GRB 190114C \citep{2019ATel12390....1M}. Therefore,  a different leptonic radiation mechanism of synchrotron radiation such as synchrotron self-Compton (SSC) in the forward shock scenario has to be considered \cite[e.g. see,][]{2004IJMPA..19.2385Z, 2015PhR...561....1K}.\\

In this study, we present a general analytic model (based on external forward shocks) where the synchrotron and SSC emission from material that is being decelerated in an arbitrary direction with respect to the observer, are obtained. Specifically, we calculate the flux from material that is in the relativistic phase and also material that is in the laterally expanding phase, and that are decelerating in either a homogeneous or a wind-like circumburst medium. As a particular case, we focus on the emission from a quasi-spherical outflow that is viewed on-axis and a relativistic top-hat jet viewed off-axis. As an application of our model, we describe the extended X-ray, optical and radio emission exhibited in GRB 170817A.  Hereafter, when we mention a jet we refer to a top-hat jet. This paper is arranged as follows: In Section 2 we construct our model and show the synchrotron and SSC light curves from material that is being decelerated in an arbitrary direction with respect to the observer. In section 3 we show the particular cases of a quasi-spherical outflow viewed on-axis and a relativistic jet viewed off-axis. In Section 4, we compare our model to the EM counterpart of GW170817. In section 5, we present our conclusions.\\

\section{Electromagnetic forward-shock emission radiated in an arbitrary direction}\label{section2}

During the fusion of a binary NS (BNS) system a wind may be ejected in practically all directions, and once the BNS merges, a relativistic jet and its correspondent quasi-spherical outflow may be powered \citep[e.g., see][]{2018MNRAS.478.4128G}. Once the jet/quasi-spherical outflow sweeps enough circumburst medium (which may have been affected by the BNS wind), the relativistic electrons accelerated through the forward shocks (FSs) are mainly cooled down by synchrotron and SSC radiation. Consequently, We derive the synchrotron and SSC fluxes in the fully adiabatic regime from material that is moving relativistically and material that is being decelerated in an arbitrary direction (with respect to the observer) by either a homogeneous or a wind-like medium. We must note that we only consider the electrons accelerated by FSs because they produce extended emissions, and not by reverse shocks as they produce short-lived emissions.

\vspace{1cm}
\subsection{Homogeneous medium}

The FS dynamics in the fully adiabatic regime for material spreading through a homogeneous medium has been explored for the case when the radiation is pointing directly in the observer's direction \cite[see, e.g.][]{1998ApJ...497L..17S},  not for the case when it points in an arbitrary direction. Therefore, we derive and present in the following section, the synchrotron and SSC fluxes during the relativistic and lateral expansion phases for radiation pointing in any arbitrary direction.

\subsubsection{Relativistic phase of the deceleration material}

\paragraph{Synchrotron radiation}

Accelerated electrons in forward-shock models are  distributed in accordance with their Lorentz factors ($\gamma_e$) and described by the electron power index $p$ as $N(\gamma_e)\,d\gamma_e \propto \gamma_e^{-p}\,d\gamma_e$ for $\gamma_m\leq \gamma_e\leq\gamma_c$, where $\gamma_m$ and $\gamma_c$ are the minimum and cooling electron Lorentz factors. These are given by:
{\small
\bary\label{ele_Lorentz_gen}
\gamma_m&=&3.1\times 10^2\,g(p)\,\varepsilon_{\rm e,-1}\Gamma_1\cr
\gamma_c&=& 5.8\times 10^4\left(\frac{1+z}{1.022}\right) (1+Y)^{-1}\varepsilon_{B,-3}^{-1}\,n_{-1}^{-1}\,\Gamma_1^{-2}\delta_{D,1}^{-1}\,t^{-1}_{\rm 1d},\,\,\,\,
\eary
}
where $g(p)=\frac{p-2}{p-1}\simeq0.17$ for $p=2.2$, $n$ is the density of the circumburst medium, $Y$ is the Compton parameter, $\varepsilon_{\rm B}$ and $\varepsilon_{\rm e}$ are the microphysical parameters associated  with the magnetic field and the energy density given to accelerate electrons, $t_{\rm 1d}$ is the timescale of the order of one day, and $\delta_D$ is the Doppler factor. The Doppler factor is  $\delta_D=\frac{1}{\Gamma(1-\mu\beta)}$ where $\mu=\cos \Delta \theta$, $\beta$ is the velocity of the material, and $\Delta \theta=\theta_{\rm obs} - \theta_{\rm j}$ is given by the viewing angle ($\theta_{\rm obs}$) and the aperture angle of the jet ($\theta_{\rm j}$). We adopt the convention $Q_{\rm x}=Q/10^{\rm x}$  in cgs units.  

Given the fact that $\epsilon^{\rm syn}_{\rm i}\propto \gamma^2_{\rm i}$ (for {\rm i=m, c} with {\rm m} the characteristic and {\rm c} the cooling break) with  eq. \ref{ele_Lorentz_gen},  the synchrotron spectral breaks ($\epsilon^{\rm syn}_{\rm m}$ and $\epsilon^{\rm syn}_{\rm c}$) and the maximum flux ($F^{\rm syn}_{\rm max}$) become:
{\small
\bary\label{ene_break_gen}
\epsilon^{\rm syn}_{\rm m}&\simeq& 15.9\,{\rm GHz}\,\, \left(\frac{1+z}{1.022}\right)^{-1}\,g(p)^2 \varepsilon^2_{e,-1}\,\varepsilon_{B,-3}^{\frac12}\,n_{-1}^{\frac{1}{2}}\, \delta_{D,1} \Gamma_1^3\,\cr
\epsilon^{\rm syn}_{\rm c}&\simeq&  6.2\,{\rm eV}  \left(\frac{1+z}{1.022}\right) (1+Y)^{-2}\, \varepsilon_{B,-3}^{-\frac32}\,n_{-1}^{-\frac{3}{2}}\,\delta_{D,1}^{-1}\Gamma_1^{-3}\,t^{-2}_{\rm 1d}\cr
F^{\rm syn}_{\rm max} &\simeq& 5.8\times 10^4\,{\rm mJy}\left(\frac{1+z}{1.022}\right)^{-1}\varepsilon_{B,-3}^{\frac12}\,n_{-1}^{\frac{3}{2}}\, D^{-2}_{26.5}\,\delta_{D,1}^6 \Gamma_1^{4}\,t^{3}_{\rm 1d}.\,\,\,\,\,.
\eary
}

Given the spectral breaks and the maximum flux from eq. \ref{ene_break_gen}, the synchrotron light curve for the fast-cooling regime ($\epsilon^{\rm syn}_{\rm c}<\epsilon^{\rm syn}_{\rm m}$) is:
{\small
\begin{eqnarray}
\label{FC_syn_bb_g}
F^{\rm syn}_{\nu}\propto\cases{
\delta_D^{\frac{19}{3}}\, \Gamma^{5}\,t^{\frac{11}{3}}\, \epsilon_\gamma^{\frac13}\,,\hspace{1.3cm} \epsilon_\gamma<\epsilon^{\rm syn}_{\rm c},\cr
\delta_D^{\frac{11}{2}}\, \Gamma^{\frac{5}{2}}\,t^{2}\,    \,\epsilon_\gamma^{-\frac{1}{2}}\,, \hspace{1.2cm}    \epsilon^{\rm syn}_{\rm c}<\epsilon_\gamma<\epsilon^{\rm syn}_{\rm m},\,\,\,\,\,\cr
\delta_D^{\frac{p+10}{2}}\, \Gamma^{\frac{3p+2}{2}}\,t^{2}\, \epsilon_\gamma^{-\frac{p}{2}},\, \hspace{0.5cm}  \epsilon^{\rm syn}_{\rm m}<\epsilon_\gamma\,. \cr
}
\end{eqnarray}
}

Meanwhile, the light curve for the slow-cooling regime ($\epsilon^{\rm syn}_{\rm m}<\epsilon^{\rm syn}_{\rm c}$) is:
{\small
\begin{eqnarray}
\label{SC_syn_bb_g}
F^{\rm syn}_{\nu}\propto\cases{
\delta_D^{\frac{17}{3}}\, \Gamma^{3}\,t^{3}\, \epsilon_\gamma^{\frac13}\,,\hspace{1.8cm} \epsilon_\gamma<\epsilon^{\rm syn}_{\rm m},\cr
\delta_D^{\frac{p+11}{2}}\, \Gamma^{\frac{3p+5}{2}}\,t^{3}\,    \,\epsilon_\gamma^{-\frac{p-1}{2}}\,, \hspace{0.4cm}    \epsilon^{\rm syn}_{\rm m}<\epsilon_\gamma<\epsilon^{\rm syn}_{\rm c},\,\,\,\,\,\cr
\delta_D^{\frac{p+10}{2}}\, \Gamma^{\frac{3p+2}{2}}\,t^{2}\, \epsilon_\gamma^{-\frac{p}{2}},\, \hspace{0.8cm}  \epsilon^{\rm syn}_{\rm c}<\epsilon_\gamma\,, \cr
}
\end{eqnarray}
}
with $\epsilon_\gamma$, in general, the energy at which the flux is observed.

\paragraph{SSC emission} 

Accelerated electrons in the FSs can up-scatter synchrotron photons from low to high energies as $\epsilon^{\rm ssc}_{\rm i}\sim\gamma^2_{\rm i} \epsilon^{\rm syn}_{\rm i}$ with a SSC flux $F^{\rm ssc}_{\rm max}\sim\, 4g(p)^{-1} \tau\,F^{\rm syn}_{\rm max}$, where {\small $\tau$} is the optical depth ($\tau=\frac13 \sigma_T n R$ with $R$ the deceleration radius, and $\sigma_T$ the Thompson cross section). Using eqs. (\ref{ele_Lorentz_gen}) and (\ref{ene_break_gen}), the spectral breaks and the maximum flux of SSC emission are (respectively):
{\small
\bary\label{energies_break_ssc}
\epsilon^{\rm ssc}_{\rm m}&\simeq& 6.2\,{\rm eV}\,\, \left(\frac{1+z}{1.022}\right)^{-1}\,g(p)^4\,\varepsilon^4_{e,-1}\,\varepsilon_{B,-3}^{\frac12}\,n_{-1}^{\frac{1}{2}}\, \delta_{D,1} \Gamma_1^5\,\cr
\epsilon^{\rm ssc}_{\rm c}&\simeq&  20.6\,{\rm GeV}  \left(\frac{1+z}{1.022}\right)^3 (1+Y)^{-4}\, \varepsilon_{B,-3}^{-\frac72}\,n_{-1}^{-\frac{7}{2}}\,\delta_{D,1}^{-3}\Gamma_1^{-7}\,\cr
&& \hspace{6.2cm}\,t^{-4}_{\rm 1d}\cr
F^{\rm ssc}_{\rm max} &\simeq& 7.8\times 10^{-3}\,{\rm mJy}\,\, \left(\frac{1+z}{1.022}\right)^{-2}\,g(p)^{-1}\varepsilon_{B,-3}^{\frac12}\,n_{-1}^{\frac{5}{2}}\, D^{-2}_{26.5}\,\cr
&& \hspace{5.0cm}\, \times\, \delta_{D,1}^7\,    \Gamma_1^{5}\,t^{4}_{\rm 1d}\,.
\eary
}
The Klein-Nishina (KN) suppression effect must be considered in SSC emission. The break energy in the KN regime is:
\be
\epsilon^{\rm ssc}_{\rm KN}\simeq 288.2\, {\rm GeV}\,(1+Y)^{-1}\, \varepsilon_{B,-1}^{-1}\,n_{-1}^{-1}\,\Gamma_1^{-2}\,t^{-1}_{\rm 1d}.
\ee
Using the synchrotron fluxes found in eqs. \ref{FC_syn_bb_g} and \ref{SC_syn_bb_g}, the SSC light curves for the fast- and slow-cooling regimes are (respectively): 
{\small
\begin{eqnarray}
\label{FC_ssc_bb_g}
F^{\rm ssc}_{\nu}\propto\cases{
\delta_D^{8}\, \Gamma^{\frac{22}{3}}\,t^{\frac{16}{3}}\, \epsilon_\gamma^{\frac13}\,,\hspace{1.3cm} \epsilon_\gamma<\epsilon^{\rm ssc}_{\rm c},\cr
\delta_D^{\frac{11}{2}}\, \Gamma^{\frac{3}{2}}\,t^{2}\,    \,\epsilon_\gamma^{-\frac{1}{2}}\,, \hspace{1.1cm}    \epsilon^{\rm ssc}_{\rm c}<\epsilon_\gamma<\epsilon^{\rm ssc}_{\rm m},\,\,\,\,\,\cr
\delta_D^{\frac{p+10}{2}}\, \Gamma^{\frac{5p-2}{2}}\,t^{2}\, \epsilon_\gamma^{-\frac{p}{2}},\, \hspace{0.5cm}  \epsilon^{\rm ssc}_{\rm m}<\epsilon_\gamma\,, \cr
}
\end{eqnarray}
}

{\small
\begin{eqnarray}
\label{SC_ssc_bb_g}
F^{\rm ssc}_{\nu}\propto\cases{
\delta_D^{\frac{20}{3}}\, \Gamma^{\frac{10}{3}}\,t^{4}\, \epsilon_\gamma^{\frac13}\,,\hspace{1.8cm} \epsilon_\gamma<\epsilon^{\rm ssc}_{\rm m},\cr
\delta_D^{\frac{p+13}{2}}\, \Gamma^{\frac{5(p+1)}{2}}\,t^{4}\,    \,\epsilon_\gamma^{-\frac{p-1}{2}}\,, \hspace{0.4cm}    \epsilon^{\rm ssc}_{\rm m}<\epsilon_\gamma<\epsilon^{\rm ssc}_{\rm c},\,\,\,\,\,\cr
\delta_D^{\frac{p+10}{2}}\, \Gamma^{\frac{5p-2}{2}}\,t^{2}\, \epsilon_\gamma^{-\frac{p}{2}},\, \hspace{1.0cm}  \epsilon^{\rm ssc}_{\rm c}<\epsilon_\gamma\,. \cr
}
\end{eqnarray}
}
\subsubsection{Lateral expansion phase of the deceleration material}
\label{lateral_exp}

As the relativistic material sweeps through the medium and decelerates, its beaming cone broadens until it reaches the field of view of the observer \citep[$\Gamma \sim \Delta \theta^{-1}$;][]{2002ApJ...570L..61G,  2017arXiv171006421G}. We will refer to this phase as the lateral expansion phase. In this phase, the Doppler factor becomes $\delta_D\approx 2\Gamma$ and the maximum flux must be corrected by dividing by $\Omega=4\pi \delta_D^2$ \citep{1986rpa..book.....R, 2018MNRAS.481.2581L, 2019ApJ...871..200F}. Taking into account that the lateral expansion phase is expected in a timescale that goes from several hours to days \citep[e.g., see][]{2015PhR...561....1K}, the fast-cooling regime for the synchrotron and SSC emission would be negligible and the slow-cooling regime will dominate. For this reason, we only derive the synchrotron and SSC light curves in the slow-cooling regime during this phase.

\paragraph{Synchrotron radiation.} 

The synchrotron spectral breaks can be calculated through eq. (\ref{ene_break_gen}) with $\delta_D\approx 2\Gamma$, and the correction of $\Omega=4\pi \delta_D^2$ for the maximum flux. In this case, the synchrotron flux for the slow-cooling regime becomes:
{\small
\begin{eqnarray}
\label{FC_syn_bb}
F^{\rm syn}_{\nu}\propto\cases{
\Gamma^{\frac{20}{3}}\,t^{3}\, \epsilon_\gamma^{\frac13}\,,\hspace{1.5cm} \epsilon_\gamma<\epsilon^{\rm syn}_{\rm m},\cr
\Gamma^{2(p+3)}\,t^{3}\,      \,\epsilon_\gamma^{-\frac{p-1}{2}}\,, \hspace{0.5cm}    \epsilon^{\rm syn}_{\rm m}<\epsilon_\gamma<\epsilon^{\rm syn}_{\rm c},\,\,\,\,\,\cr
\Gamma^{2(p+2)}\,t^{2}\,   \, \epsilon_\gamma^{-\frac{p}{2}},\, \hspace{0.8cm}  \epsilon^{\rm syn}_{\rm c}<\epsilon_\gamma\,. \cr
}
\end{eqnarray}
}

\paragraph{SSC emission.} 

The SSC spectral breaks can be calculated through eq. (\ref{energies_break_ssc})  with the same corrections of $\delta_D$ and $\Omega$ as for the synchrotron radiation.  In this case, the SSC flux for the slow-cooling regime becomes:
{\small
\begin{eqnarray}
\label{FC_syn_bb}
F^{\rm ssc}_{\nu}\propto\cases{
\Gamma^8\,t^{4}\, \epsilon_\gamma^{\frac13}\,,\hspace{1.45cm} \epsilon_\gamma<\epsilon^{\rm ssc}_{\rm m},\cr
\Gamma^{3p+7}\,t^{4}\,    \,\epsilon_\gamma^{-\frac{p-1}{2}}\,, \hspace{0.4cm}    \epsilon^{\rm ssc}_{\rm m}<\epsilon_\gamma<\epsilon^{\rm ssc}_{\rm c},\,\,\,\,\,\cr
\Gamma^{3p+2}\,t^{2}\, \epsilon_\gamma^{-\frac{p}{2}},\, \hspace{0.8cm}  \epsilon^{\rm ssc}_{\rm c}<\epsilon_\gamma\,. \cr
}
\end{eqnarray}
}

\subsection{Wind-like medium}
\label{modelo: wind}

A non-homogeneous density produced by the ejected mass near the vicinity of the NS binary system has been studied numerically  \citep{2014ApJ...784L..28N, 2013ApJ...778L..16H, 2013ApJ...773...78B}.  \cite{2014ApJ...784L..28N} studied the propagation of a relativistic jet through the ejected mass using a density profile of $\rho(r)\propto \frac{M_{\rm ej}}{r_0^3} r^{-k}$ with $r_0$ the initial radius and $M_{\rm ej}$ the ejecta mass. Since the density profile around the merger can be approximated as a medium with ${\rm k= 2}$, we derive the dynamics for material either in the relativistic or the lateral expanding phase in a density profile which scales as $\rho(r)\propto r^{-2}$. Specifically, we assume that the wind-like medium is $\rho(r)=A r^{-2}$, with $A=\frac{\dot{M}}{4\pi\, v}=A_\star\,5\times 10^{11}\,{\rm g\,cm^{-1}}$, $\dot{M}$ the mass-loss rate, $v$ the velocity of the outflow, and $A_\star$ a dimensional parameter ($A_\star \sim$10$^{-1}-$10$^{-6}$). Given that the lateral expansion phase is expected in a timescale that goes from several hours to hundreds of days \citep[e.g., see][]{2015PhR...561....1K}, thus the lateral expansion phase in a wind-like medium is negligible (compared to the relativistic phase). Hence, we only derive the synchrotron and SSC fluxes of the relativistic phase.

\subsubsection{Relativistic phase of the deceleration material}

\paragraph{Synchrotron emission.}

The minimum and cooling Lorentz factors in a wind-like medium are given by:
{\small
\bary\label{ele_Lorentz_gen_w}
\gamma_m&=& 3.1\times 10^2\,g(p)\,\varepsilon_{e,-1}\Gamma_1\cr
\gamma_c&=&  1.4\times 10^3\left(\frac{1+z}{1.022}\right)^{-1}\xi^{2} (1+Y)^{-1}\varepsilon_{B,-3}^{-1}\,A_{\star,-4}^{-1}\,\delta_{D,1}\,\cr
&&\hspace{5.7cm}\times t_{\rm 10s},\hspace{0.4cm}
\eary
}
where $\xi$ is a constant parameter ($\xi \approx$1) \citep{1998ApJ...493L..31P}. 

Using eq. (\ref{ele_Lorentz_gen_w}) in the synchrotron emission, the spectral breaks and the maximum flux are:
{\small
\bary\label{energies_break_gen_w}
\epsilon^{\rm syn}_{\rm m}&\simeq& 9.4\times 10^{4}\,{\rm GHz}\,g(p)^2\,\xi^{-2}\varepsilon^2_{e,-1}\,\varepsilon_{B,-3}^{\frac12}\,A_{\star,-4}^{\frac{1}{2}}\, \Gamma_1^2\,t^{-1}_{\rm 10s}\,\cr
\epsilon^{\rm syn}_{\rm c}&\simeq&  1.1\,{\rm eV}  \left(\frac{1+z}{1.022}\right)^{-2}\xi^{2} (1+Y)^{-2}\, \varepsilon_{B,-3}^{-\frac32}\,A_{\star,-4}^{-\frac{3}{2}}\,\delta_{D,1}^{2}\,t_{\rm 10s}\cr
F^{\rm syn}_{\rm max} &\simeq& 18.4\,{\rm mJy}\,\, \left(\frac{1+z}{1.022}\right)^{2}\,\varepsilon_{B,-3}^{\frac12}\,A_{\star,-4}^{\frac{3}{2}}\, D^{-2}_{26.5}\,\delta_{D,1}^3 \Gamma_1.
\eary
}

Using eq. (\ref{energies_break_gen_w}) in the synchrotron emission, the flux for the fast- and slow-cooling regimes are (respectively):
{\small
\begin{eqnarray}
\label{FC_syn_bb_gen}
F^{\rm syn}_{\nu}\propto\cases{
\delta_D^{\frac{7}{3}}\, \Gamma\,t^{-\frac{1}{3}}\, \epsilon_\gamma^{\frac13}\,,\hspace{1.3cm} \epsilon_\gamma<\epsilon_{\rm c},\cr
\delta_D^4\, \Gamma\,t^{\frac12}\,    \,\epsilon_\gamma^{-\frac{1}{2}}\,, \hspace{1.2cm}    \epsilon_{\rm c}<\epsilon_\gamma<\epsilon_{\rm m},\,\,\,\,\,\cr
\delta_D^4\, \Gamma^p\,t^{-\frac{p}{2}+1}\, \epsilon_\gamma^{-\frac{p}{2}},\, \hspace{0.6cm}  \epsilon_{\rm m}<\epsilon_\gamma\,, \cr
}
\end{eqnarray}
}

{\small
\begin{eqnarray}
\label{SC_syn_bb_gen}
F^{\rm syn}_{\nu}\propto\cases{
\delta_D^{3}\, \Gamma^{\frac13}\,t^{\frac13}\, \epsilon_\gamma^{\frac13}\,,\hspace{1.4cm} \epsilon_\gamma<\epsilon_{\rm m},\cr
\delta_D^{3}\, \Gamma^{p}\,t^{-\frac{p-1}{2}}\,    \,\epsilon_\gamma^{-\frac{p-1}{2}}\,, \hspace{0.4cm}    \epsilon_{\rm m}<\epsilon_\gamma<\epsilon_{\rm c},\,\,\,\,\,\cr
\delta_D^{4}\, \Gamma^{p}\,t^{-\frac{p}{2}+1}\, \epsilon_\gamma^{-\frac{p}{2}},\, \hspace{0.8cm}  \epsilon_{\rm c}<\epsilon_\gamma\,. \cr
}
\end{eqnarray}
}

\paragraph{SSC emission} 

Given the electron Lorentz factors (eq. \ref{ele_Lorentz_gen_w}) and the synchrotron spectral breaks (eq. \ref{energies_break_gen_w}), the SSC spectral breaks and the maximum flux are:

{\small
\bary\label{energies_break_gen_l_w}
\epsilon^{\rm ssc}_{\rm m}&\simeq& 3.6\,{\rm keV}\,\,g(p)^{4}\xi^{-2}\varepsilon^4_{e,-1}\,\epsilon_{B,-3}^{\frac12}\,A_{\star,-4}^{\frac{1}{2}}\, \Gamma_1^4\,t^{-1}_{\rm 10s}\,\cr
\epsilon^{\rm ssc}_{\rm c}&\simeq&  2.3\,{\rm MeV}  \left(\frac{1+z}{1.022}\right)^{-4} \xi^{6}(1+Y)^{-4}\, \epsilon_{B,-3}^{-\frac72}\,A_{\star,-4}^{-\frac{7}{2}}\,\delta_{D,1}^{4}\,t^3_{\rm 10s}\cr
F^{\rm ssc}_{\rm max} &\simeq& 1.1\times 10^{-4}\,{\rm mJy}\,\, \left(\frac{1+z}{1.022}\right)^3\,g(p)^{-1}\xi^{-2}\varepsilon_{B,-3}^{\frac12}\,A_{\star,-4}^{\frac{5}{2}}\, D^{-2}_{26.5}\,\cr
&&\hspace{5.3cm}\times\,\,\delta_{D,1}^2 t^{-1}_{\rm 10s}.
\eary
}
The break energy in the SSC emission due to the KN effect is:
\be
\epsilon^{\rm ssc}_{\rm KN}\simeq 7.2\,{\rm GeV} \left(\frac{1+z}{1.022}\right)^{-2}\,(1+Y)^{-1}\, 
\xi^2\, \varepsilon_{B,-3}^{-1}\,A^{-1}_{\star,-4}\,\delta_{D,1}^{2}\,.
\ee
Using the SSC spectrum, the spectral breaks and the maximum flux (eq. \ref{energies_break_gen_l_w}), the SSC light curves for fast- and slow-cooling regimes are (respectively):
{\small
\begin{eqnarray}
\label{FC_ssc_bb_gen}
F^{\rm ssc}_{\nu}\propto\cases{
\delta_D^{\frac{2}{3}}\,t^{-2}\, \epsilon_\gamma^{\frac13}\,,\hspace{1.9cm} \epsilon_\gamma<\epsilon_{\rm c},\cr
\delta_D^4\, t^{\frac12}\,    \,\epsilon_\gamma^{-\frac{1}{2}}\,, \hspace{1.85cm}    \epsilon_{\rm c}<\epsilon_\gamma<\epsilon_{\rm m},\,\,\,\,\,\cr
\delta_D^4\, \Gamma^{2p-2}\,t^{-\frac{p}{2}+1}\, \epsilon_\gamma^{-\frac{p}{2}},\, \hspace{0.5cm}  \epsilon_{\rm m}<\epsilon_\gamma\,, \cr
}
\end{eqnarray}
}

{\small
\begin{eqnarray}
\label{SC_ssc_bb_gen}
F^{\rm ssc}_{\nu}\propto\cases{
\delta_D^{2}\, \Gamma^{-\frac43}\,t^{-\frac23}\, \epsilon_\gamma^{\frac13}\,,\hspace{1.8cm} \epsilon_\gamma<\epsilon_{\rm m},\cr
\delta_D^{2}\, \Gamma^{2(p-1)}\,t^{-\frac{p+1}{2}}\,    \,\epsilon_\gamma^{-\frac{p-1}{2}}\,, \hspace{0.5cm}    \epsilon_{\rm m}<\epsilon_\gamma<\epsilon_{\rm c},\,\,\,\,\,\cr
\delta_D^{4}\, \Gamma^{2p-2}\,t^{-\frac{p}{2}+1}\, \epsilon_\gamma^{-\frac{p}{2}},\, \hspace{1.1cm}  \epsilon_{\rm c}<\epsilon_\gamma\,. \cr
}
\end{eqnarray}
}
\section{Quasi-spherical outflow and off-axis jet}
\begin{figure}
{ \centering
\resizebox*{0.55\textwidth}{0.4\textheight}
{\includegraphics{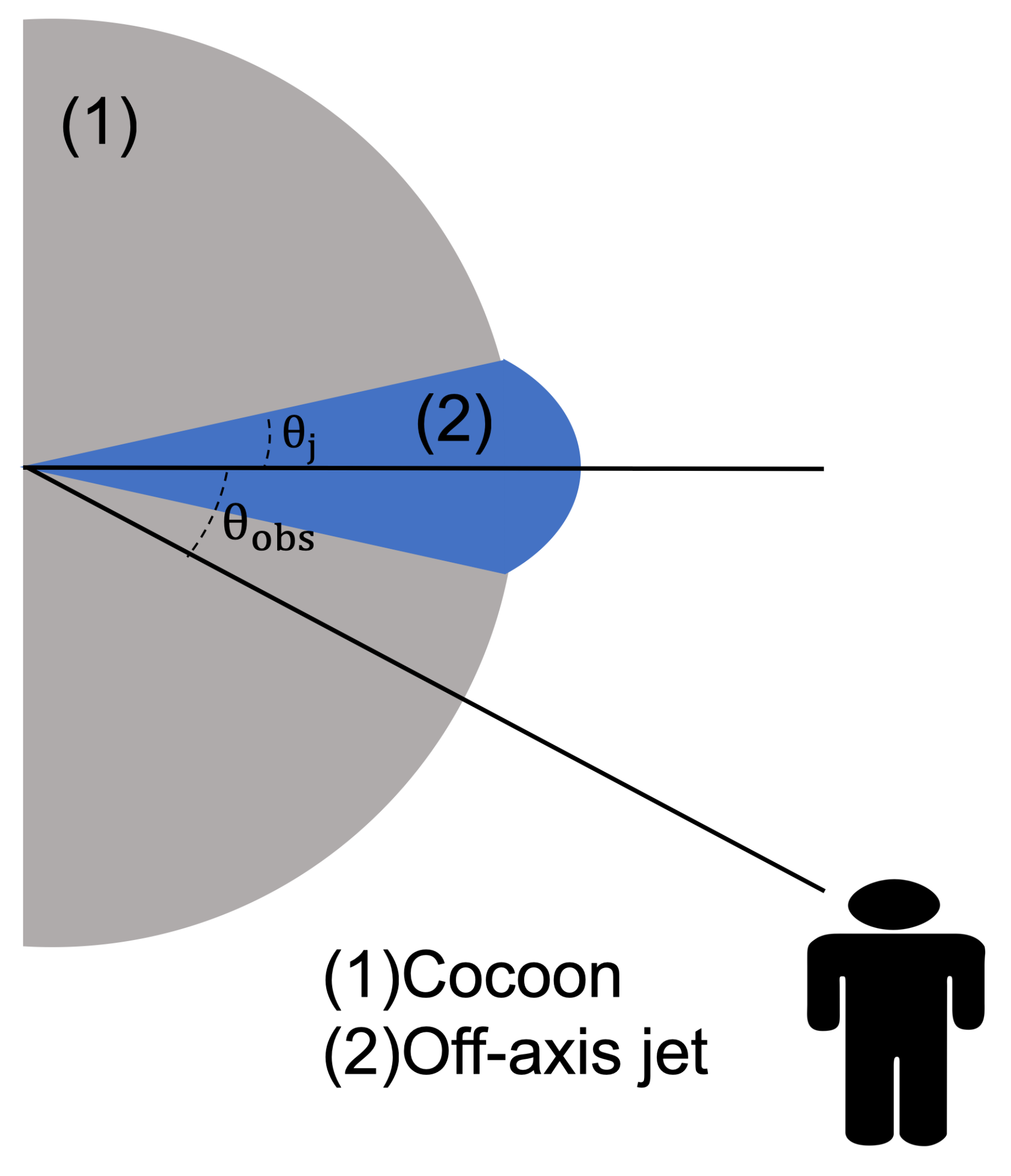}}
}
\caption{Schematic representation of the quasi-spherical outflow, the off-axis jet, and the observer. The quasi-spherical outflow emits photons at nearly all the viewing angles while the off-axis jet emits mainly towards the propagation direction.}
\label{fig1:sketch}
\end{figure}

We now calculate the electromagnetic radiation for the specific case of an off-axis jet plus a wide-angle quasi-spherical outflow in a homogeneous and wind-like medium. Figure \ref{fig1:sketch} shows the schematic representation of the electromagnetic emission produced by an off-axis jet and its corresponding quasi-spherical outflow. On one hand, the decelerated material from the quasi-spherical outflow releases photons at nearly all the viewing angles. On the other hand, only a small fraction of the emission from the material in the relativistic jet is observable (due to relativistic beaming $\theta_{\rm{j}} \propto 1 / \Gamma$). Once the outflow and the off-axis jet sweep up enough circumburst medium, the electron population inside either case cool down emitting synchrotron and SSC radiation. In the case of the quasi-spherical outflow, the Doppler factor can be approximated to $\delta_D\approx 2\Gamma$ and in the case of the off-axis jet, this factor can be approximated to  $\delta_D\simeq \frac{2}{\Gamma\Delta\theta^2}$ for  $\Gamma^2 \Delta \theta^2\gg1$. With these approximations in the obtained fluxes from Section \ref{section2} we derive the dynamics for each case.

\subsection{Quasi-spherical outflow}

The synchrotron and SSC emission of a quasi-spherical outflow in the relativistic phase and in the lateral expansion phase, moving through either a homogeneous medium or a wind-like medium, are next calculated.

\subsubsection{Homogeneous Medium: Relativistic phase}
Assuming that the equivalent kinetic energy ($E_{\rm k}$) associated to the material that is accelerated in the quasi-spherical outflow can be written as $E_{\rm k}=\tilde{E}\,\Gamma^{-\alpha_s}$, and considering the Blandford-McKee solution \citep{1976PhFl...19.1130B},
where $E_{\rm k}=\frac{4\pi}{3}\,m_p\,n\,R^3\Gamma^2$, then the bulk Lorentz factor ($\Gamma$) can be written as:
\be\label{Gamma_c}
\Gamma= 3.8\,\left(\frac{1+z}{1.022}\right)^{\frac{3}{\alpha_s+8}}\,n_{-1}^{-\frac{1}{\alpha_s+8}}\,\tilde{E}^{\frac{1}{\alpha_s+8}}_{50} t_{\rm 1d}^{-\frac{3}{\alpha_s+8}}\,, 
\ee
where ${\rm z}$ is the redshift, $\tilde{E}$ is the fiducial energy given by $\tilde{E}=\frac{32\pi}{3}\,m_p(1+z)^{-3}\,n\Gamma^{\alpha_s+8}\,t^3$,  $m_p$ is the proton mass and $\alpha_s$ is the power index of the velocity distribution. Using eq. (\ref{Gamma_c}) in eqs. (\ref{FC_syn_bb_g}), (\ref{SC_syn_bb_g}), (\ref{FC_ssc_bb_g}) and (\ref{SC_ssc_bb_g}), we next obtain the synchrotron and SSC emission of a quasi-spherical outflow in the relativistic phase moving through a homogeneous medium.

\paragraph{Synchrotron light curves} Given the evolution of the bulk Lorentz factor (eq. \ref{Gamma_c}) and the synchrotron spectra (eqs. \ref{FC_syn_bb_gen} and \ref{SC_syn_bb_gen}), the fluxes for the fast-cooling and slow-cooling regimes can be written as (respectively):
{\small
\begin{eqnarray}
\label{FC_syn_bb_c}
F^{\rm syn}_{\nu}\propto\cases{
t^{\frac{11\alpha_s +4}{3(\alpha_s + 8)}}\, \epsilon_\gamma^{\frac13}\,,\hspace{1.3cm} \epsilon_\gamma<\epsilon_{\rm c},\cr
t^{\frac{2(\alpha_s - 1)}{\alpha_s + 8}}\,    \,\epsilon_\gamma^{-\frac{1}{2}}\,, \hspace{1.2cm}    \epsilon_{\rm c}<\epsilon_\gamma<\epsilon_{\rm m},\,\,\,\,\,\cr
t^{\frac{2(\alpha_s-3p+2)}{\alpha_s + 8}}\, \epsilon_\gamma^{-\frac{p}{2}},\, \hspace{0.6cm}  \epsilon_{\rm m}<\epsilon_\gamma\,, \cr
}
\end{eqnarray}
}

{\small
\begin{eqnarray}
\label{SC_syn_bb_c}
F^{\rm syn}_{\nu}\propto\cases{
t^{\frac{3\alpha_s + 4}{\alpha_s + 8}}\, \epsilon_\gamma^{\frac13}\,,\hspace{2.0cm} \epsilon_\gamma<\epsilon^{\rm syn}_{\rm m},\cr
t^{\frac{3(\alpha_s-2p+2)}{\alpha_s + 8}}\,    \,\epsilon_\gamma^{-\frac{p-1}{2}}\,, \hspace{0.8cm}    \epsilon^{\rm syn}_{\rm m}<\epsilon_\gamma<\epsilon^{\rm syn}_{\rm c},\,\,\,\,\,\cr
t^{\frac{2(\alpha_s-3p+2)}{\alpha_s + 8}}\, \epsilon_\gamma^{-\frac{p}{2}},\, \hspace{1.2cm}  \epsilon^{\rm syn}_{\rm c}<\epsilon_\gamma\,. \cr
}
\end{eqnarray}
}

The synchrotron spectral breaks $\epsilon^{\rm syn}_{\rm m}$ and $\epsilon^{\rm syn}_{\rm c}$, and the maximum flux are given in eq. \ref{energies_break_syn_c_h}.

\paragraph{SSC light curves}  Given the evolution of the bulk Lorentz factor (eq. \ref{Gamma_c}) and the SSC spectra (eqs. \ref{FC_ssc_bb_gen} and \ref{SC_ssc_bb_gen}), the fluxes for the fast-cooling and slow cooling regimes can be written as (respectively):
{\small
\begin{eqnarray}
\label{FC_ssc_bb_c}
F^{\rm ssc}_{\nu}\propto\cases{
t^{\frac{8(2\alpha_s +1)}{3(\alpha_s + 8)}}\, \epsilon_\gamma^{\frac13}\,,\hspace{1.0cm} \epsilon_\gamma<\epsilon^{\rm ssc}_{\rm c},\cr
t^{\frac{2\alpha_s +1}{\alpha_s + 8}}\,    \,\epsilon_\gamma^{-\frac{1}{2}}\,, \hspace{1.0cm}    \epsilon^{\rm ssc}_{\rm c}<\epsilon_\gamma<\epsilon^{\rm ssc}_{\rm m},\,\,\,\,\,\cr
t^{\frac{2\alpha_s - 9p+10}{\alpha_s + 8}}\, \epsilon_\gamma^{-\frac{p}{2}},\, \hspace{0.5cm}  \epsilon^{\rm ssc}_{\rm m}<\epsilon_\gamma\,, \cr
}
\end{eqnarray}
}

{\small
\begin{eqnarray}
\label{FC_ssc_bb_c}
F^{\rm ssc}_{\nu}\propto\cases{
t^{\frac{4(\alpha_s + 2)}{\alpha_s + 8}}\, \epsilon_\gamma^{\frac13}\,,\hspace{1.8cm} \epsilon_\gamma<\epsilon^{\rm ssc}_{\rm m},\cr
t^{\frac{4\alpha_s -9p +11}{\alpha_s + 8}}\,    \,\epsilon_\gamma^{-\frac{p-1}{2}}\,, \hspace{0.8cm}    \epsilon^{\rm ssc}_{\rm m}<\epsilon_\gamma<\epsilon^{\rm ssc}_{\rm c},\,\,\,\,\,\cr
t^{\frac{2\alpha_s - 9p+10}{\alpha_s + 8}}\, \epsilon_\gamma^{-\frac{p}{2}},\, \hspace{1.15cm}  \epsilon^{\rm ssc}_{\rm c}<\epsilon_\gamma\,. \cr
}
\end{eqnarray}
}
The SSC spectral breaks $\epsilon^{\rm ssc}_{\rm m}$ and $\epsilon^{\rm ssc}_{\rm c}$, and the maximum flux are given in eq.  (\ref{energies_break_ssc_c_h})

\subsubsection{Homogeneous Medium: Lateral expansion phase}

The relativistic beaming effect of the emitting shock causes the afterglow emission to be beamed into a beaming angle ($\theta_b \sim 1/\Gamma$), which for fast flows (v$\sim$c) is narrower than the angle with which the jet is launched ($\theta_l$) and narrower than the observers viewing angle. Eventually, when the outflow decelerates, the  beaming effect weakens and the emission inside the beaming cone expands sideways and broadens. When $\theta_b \geq \theta_l$, a break in the light curve is expected \citep[e.g., see][]{2003ApJ...592.1002S, 2017ApJ...850L..24G}.\\

Considering the Blandford-McKee solution and the fact that $E_{\rm k}=\tilde{E}\,\Gamma^{-\alpha_s}$ during the lateral expansion phase, the bulk Lorentz factor can be written as

\be\label{Gamma_c_l}
\Gamma= 2.1\left(\frac{1+z}{1.022}\right)^{\frac{3}{\alpha_s+6}}\,n_{-1}^{-\frac{1}{\alpha_s+6}}\,\beta^{-\frac{\alpha_s}{\alpha_s+6}}\tilde{E}^{\frac{1}{\alpha_s+6}}_{50} t_{\rm 30d}^{-\frac{3}{\alpha_s+6}}\,. 
\ee

Using eq. (\ref{Gamma_c_l}) in eqs. (\ref{FC_syn_bb_g}), (\ref{SC_syn_bb_g}), (\ref{FC_ssc_bb_g}) and (\ref{SC_ssc_bb_g}), we next obtain the synchrotron and SSC of a quasi-spherical outflow in the lateral expansion phase moving through a homogeneous medium.

\paragraph{Synchrotron light curves} The synchrotron light curve for the slow-cooling regime is:

{\small
\begin{eqnarray}
\label{FC_syn_bb}
F^{\rm syn}_{\nu}\propto\cases{
t^{\frac{3\alpha_s -2}{\alpha_s + 6}}\, \epsilon_\gamma^{\frac13}\,,\hspace{1.7cm} \epsilon_\gamma<\epsilon_{\rm m},\cr
t^{\frac{3(\alpha_s -2p)}{\alpha_s + 6}}\,    \,\epsilon_\gamma^{-\frac{p-1}{2}}\,, \hspace{0.8cm}    \epsilon_{\rm m}<\epsilon_\gamma<\epsilon_{\rm c},\,\,\,\,\,\cr
t^{\frac{2(\alpha_s - 3p)}{\alpha_s + 6}}\, \epsilon_\gamma^{-\frac{p}{2}},\, \hspace{1.2cm}  \epsilon_{\rm c}<\epsilon_\gamma\,. \cr
}
\end{eqnarray}
}
Once the bulk Lorentz factor becomes less than $\sim 2$, the quasi-spherical outflow goes into a non-relativistic phase. In this case, the spectral breaks and the maximum flux evolve as: {\small $\epsilon_{\rm m}\propto t^{-\frac{15}{\alpha_s+5}}$, $\epsilon_{\rm c}\propto t^{-\frac{2\alpha_s+1}{\alpha_s+5}}$ and $F_{\rm max}\propto t^{\frac{3(\alpha_s+1)}{\alpha_s+5}}$}.  The synchrotron light curve in the non-relativistic phase is 

{\small
\begin{eqnarray}
\label{FC_syn_bb}
F^{\rm syn}_{\nu}\propto\cases{
t^{\frac{3\alpha_s+8}{\alpha_s + 5}}\, \epsilon_\gamma^{\frac13}\,,\hspace{1.7cm} \epsilon_\gamma<\epsilon_{\rm m},\cr
t^{\frac{6\alpha_s - 15p + 21}{2(\alpha_s + 5)}}\,    \,\epsilon_\gamma^{-\frac{p-1}{2}}\,, \hspace{0.8cm}    \epsilon_{\rm m}<\epsilon_\gamma<\epsilon_{\rm c},\,\,\,\,\,\cr
t^{\frac{4\alpha_s - 15p+20}{2(\alpha_s + 5)}}\, \epsilon_\gamma^{-\frac{p}{2}},\, \hspace{1.2cm}  \epsilon_{\rm c}<\epsilon_\gamma\,. \cr
}
\end{eqnarray}
}

\paragraph{SSC light curves} The SSC light curve for the slow-cooling regime is:

{\small
\begin{eqnarray}
\label{FC_syn_bb}
F^{\rm ssc}_{\nu}\propto\cases{
t^{\frac{4\alpha_s}{\alpha_s + 6}}\, \epsilon_\gamma^{\frac13}\,,\hspace{1.95cm} \epsilon_\gamma<\epsilon^{\rm ssc}_{\rm m},\cr
t^{\frac{4\alpha_s -9p +3}{\alpha_s + 6}}\,    \,\epsilon_\gamma^{-\frac{p-1}{2}}\,, \hspace{0.8cm}    \epsilon^{\rm ssc}_{\rm m}<\epsilon_\gamma<\epsilon^{\rm ssc}_{\rm c},\,\,\,\,\,\cr
t^{\frac{2\alpha_s - 9p+6}{\alpha_s + 6}}\, \epsilon_\gamma^{-\frac{p}{2}},\, \hspace{1.2cm}  \epsilon^{\rm ssc}_{\rm c}<\epsilon_\gamma\,. \cr
}
\end{eqnarray}
}

In the case when the quasi-spherical outflow is in the non-relativistic phase, the spectral breaks and the maximum flux evolve as: {\small $\epsilon^{\rm ssc}_{\rm m}\propto t^{-\frac{27}{\alpha_s+5}}$, $\epsilon^{\rm ssc}_{\rm c}\propto t^{-\frac{4\alpha_s}{\alpha_s+5}}$ and $F^{\rm ssc}_{\rm max}\propto t^{\frac{4(\alpha_s+5)}{\alpha_s+5}}$}.  The SSC light curve in the non-relativistic phase is 

{\small
\begin{eqnarray}
\label{FC_syn_bb}
F^{\rm ssc}_{\nu}\propto\cases{
t^{\frac{4\alpha_s+14}{\alpha_s + 5}}\, \epsilon_\gamma^{\frac13}\,,\hspace{1.95cm} \epsilon_\gamma<\epsilon^{\rm ssc}_{\rm m},\cr
t^{\frac{8\alpha_s +37- 27p}{2(\alpha_s + 5)}}\,    \,\epsilon_\gamma^{-\frac{p-1}{2}}\,, \hspace{0.8cm}    \epsilon^{\rm ssc}_{\rm m}<\epsilon_\gamma<\epsilon^{\rm ssc}_{\rm c},\,\,\,\,\,\cr
t^{\frac{4\alpha_s + 38 - 27p}{2(\alpha_s + 5}}\, \epsilon_\gamma^{-\frac{p}{2}},\, \hspace{1.2cm}  \epsilon^{\rm ssc}_{\rm c}<\epsilon_\gamma\,. \cr
}
\end{eqnarray}
}

\subsubsection{Wind-like Medium}

As already stated in Section \ref{modelo: wind}, for the wind-like medium the lateral expansion phase is negligible compared to the relativistic phase. Thus, we only calculate the synchrotron and SSC fluxes of the relativistic phase in this regime.

Considering the Blandford-McKee solution for a wind-like medium, the bulk Lorentz factor can be written as:
\be\label{Gamma_c_w}
\Gamma= 16.7\left(\frac{1+z}{1.022}\right)^{\frac{1}{\alpha_s+4}}\,\xi^{-\frac{2}{\alpha_s+4}}A_{\star,-4}^{-\frac{1}{\alpha_s+4}}\,\tilde{E}^{\frac{1}{\alpha_s+4}}_{50} t_{\rm 10s}^{-\frac{1}{\alpha_s+4}}\,,
\ee

where $\tilde{E}=\frac{32\pi}{3}\,(1+z)^{-1}\,\xi^2 A_{\star}\Gamma^{\alpha_s+4}\,t$.  Using eq. (\ref{Gamma_c_w}) in (\ref{FC_syn_bb_gen}), (\ref{SC_syn_bb_gen}), (\ref{FC_ssc_bb_gen}) and (\ref{SC_ssc_bb_gen}), we next obtain the synchrotron and SSC light curves of a quasi-spherical outflow in the relativistic phase moving through a wind-like medium.

\paragraph{Synchrotron light curves} Given the synchrotron spectrum, the light curves for the fast- and the slow-cooling regimes are (respectively): 

{\small
\begin{eqnarray}
\label{FC_syn_bb}
F^{\rm syn}_{\nu}\propto\cases{
t^{-\frac{\alpha_s + 8}{3(\alpha_s + 4)}}\, \epsilon_\gamma^{\frac13}\,,\hspace{1.8cm} \epsilon_\gamma<\epsilon^{\rm syn}_{\rm c},\cr
t^{\frac{\alpha_s - 2}{2(\alpha_s + 4)}}\,    \,\epsilon_\gamma^{-\frac{1}{2}}\,, \hspace{1.6cm}    \epsilon^{\rm syn}_{\rm c}<\epsilon_\gamma<\epsilon^{\rm syn}_{\rm m},\,\,\,\,\,\cr
t^{\frac{2\alpha_s-6p+4-\alpha_s p}{2(\alpha_s + 4)}}\, \epsilon_\gamma^{-\frac{p}{2}},\, \hspace{1.cm}  \epsilon^{\rm syn}_{\rm m}<\epsilon_\gamma\,, \cr
}
\end{eqnarray}
}

{\small
\begin{eqnarray}
\label{FC_syn_bb}
F^{\rm syn}_{\nu}\propto\cases{
t^{\frac{\alpha_s}{3(\alpha_s + 4)}}\, \epsilon_\gamma^{\frac13}\,,\hspace{1.95cm} \epsilon_\gamma<\epsilon^{\rm syn}_{\rm m},\cr
t^{\frac{\alpha_s-6p+2-\alpha_s p}{2(\alpha_s + 4)}}\,    \,\epsilon_\gamma^{-\frac{p-1}{2}}\,, \hspace{0.8cm}    \epsilon^{\rm syn}_{\rm m}<\epsilon_\gamma<\epsilon^{\rm syn}_{\rm c},\,\,\,\,\,\cr
t^{\frac{2\alpha_s-6p+4-\alpha_s p}{2(\alpha_s + 4)}}\, \epsilon_\gamma^{-\frac{p}{2}},\, \hspace{1.1cm}  \epsilon^{\rm syn}_{\rm c}<\epsilon_\gamma\,. \cr
}
\end{eqnarray}
}

The synchrotron spectral breaks $\epsilon^{\rm syn}_{\rm m}$ and $\epsilon^{\rm syn}_{\rm c}$, and the maximum flux are given in eq.  (\ref{energies_break_syn_w}).

\paragraph{SSC light curves} Given the synchrotron spectrum, the light curves for the fast- and the slow-cooling regimes are (respectively): 

{\small
\begin{eqnarray}
\label{FC_syn_bb}
F^{\rm ssc}_{\nu}\propto\cases{
t^{-\frac{2(3\alpha_s + 10)}{3(\alpha_s + 4)}}\, \epsilon_\gamma^{\frac13}\,,\hspace{1.cm} \epsilon_\gamma<\epsilon^{\rm ssc}_{\rm c},\cr
t^{\frac{\alpha_s}{2(\alpha_s + 4)}}\,    \,\epsilon_\gamma^{-\frac{1}{2}}\,, \hspace{1.2cm}    \epsilon^{\rm ssc}_{\rm c}<\epsilon_\gamma<\epsilon^{\rm ssc}_{\rm m},\,\,\,\,\,\cr
t^{\frac{2\alpha_s-8p+8-\alpha_s p}{2(\alpha_s + 4)}}\, \epsilon_\gamma^{-\frac{p}{2}},\, \hspace{0.5cm}  \epsilon^{\rm ssc}_{\rm m}<\epsilon_\gamma\,, \cr
}
\end{eqnarray}
}

{\small
\begin{eqnarray}
\label{FC_syn_bb}
F^{\rm ssc}_{\nu}\propto\cases{
t^{-\frac{2(\alpha_s + 2)}{3(\alpha_s + 4)}}\, \epsilon_\gamma^{\frac13}\,,\hspace{1.6cm} \epsilon_\gamma<\epsilon^{\rm ssc}_{\rm m},\cr
t^{-\frac{\alpha_s+8p+\alpha_s p}{2(\alpha_s + 4)}}\,    \,\epsilon_\gamma^{-\frac{p-1}{2}}\,, \hspace{0.8cm}    \epsilon^{\rm ssc}_{\rm m}<\epsilon_\gamma<\epsilon^{\rm ssc}_{\rm c},\,\,\,\,\,\cr
t^{\frac{2\alpha_s-8p+8-\alpha_s p}{2(\alpha_s + 4)}}\, \epsilon_\gamma^{-\frac{p}{2}},\, \hspace{1.0cm}  \epsilon^{\rm ssc}_{\rm c}<\epsilon_\gamma\,. \cr
}
\end{eqnarray}
}

The SSC spectral breaks $\epsilon^{\rm ssc}_{\rm m}$ and $\epsilon^{\rm ssc}_{\rm c}$, and the maximum flux are given in eq.  (\ref{energies_break_ssc_w}).

\subsection{Off-axis jet}

The synchrotron and SSC emission of an off-axis jet in the relativistic and the lateral expansion phase, moving through either a homogeneous medium or a wind-like medium, are next calculated.

\subsubsection{Homogeneous Medium: Relativistic phase}

The equivalent kinetic energy is $E_{\rm k}=\frac{\tilde{E}}{1-\cos\theta_j}\approx \frac{2\tilde{E}}{\theta^2_j}$ with the fiducial energy   $\tilde{E}=\frac{16\pi}{3}\,m_p(1+z)^{-3}\,n\,\theta^{2}_{\rm j}\,\Delta \theta^{-6} \,\Gamma^{2}\,t^3$.  In this case, the bulk Lorentz  factor evolves as:
{\small
\be\label{Gamma_o}
\Gamma=321.1\,\left(\frac{1+z}{1.022}\right)^{\frac{3}{2}}  \,n^{-\frac{1}{2}}_{-1}\,\tilde{E}^{\frac{1}{2}}_{50}\,\theta_{j,5^\circ}^{-1}\,\Delta\theta_{15^\circ}^3\,t^{-\frac{3}{2}}_{\rm 1d}\,.
\ee
}

Using eq. (\ref{Gamma_o}) in (\ref{FC_syn_bb_g}), (\ref{SC_syn_bb_g}), (\ref{FC_ssc_bb_g}) and (\ref{SC_ssc_bb_g}), we next obtain the synchrotron and SSC light curves of an off-axis jet in the relativistic phase moving through a homogeneous medium.

\paragraph{Synchrotron light curves} The synchrotron light curves for the fast- and slow-cooling regimes, are (respectively):  

{\small
\begin{eqnarray}
\label{scsyn_t}
F^{\rm syn}_{\nu}\propto\cases{
t^{\frac{17}{3}} \, \epsilon_\gamma^{\frac13},\hspace{1.85cm} \epsilon_\gamma<\epsilon^{\rm syn}_{\rm m},\cr
t^{\frac{13}{2}}\,\epsilon_\gamma^{-\frac{p-1}{2}}, \hspace{0.5cm}       \epsilon^{\rm syn}_{\rm m}<\epsilon_\gamma<\epsilon^{\rm syn}_{\rm c},\,\,\,\,\,\cr
t^{\frac{-3p+16}{2}}\,\epsilon_\gamma^{-\frac{p}{2}},  \hspace{0.95cm}   \,\epsilon^{\rm syn}_{\rm c}<\epsilon_\gamma\,, \cr
}
\end{eqnarray}
}

{\small
\begin{eqnarray}
\label{scsyn_t}
F^{\rm syn}_{\nu}\propto\cases{
t^7 \, \epsilon_\gamma^{\frac13},\hspace{1.85cm} \epsilon_\gamma<\epsilon^{\rm syn}_{\rm m},\cr
t^{\frac{-3p+15}{2}}\,\epsilon_\gamma^{-\frac{p-1}{2}}, \hspace{0.5cm}       \epsilon^{\rm syn}_{\rm m}<\epsilon_\gamma<\epsilon^{\rm syn}_{\rm c},\,\,\,\,\,\cr
t^{\frac{-3p+16}{2}}\,\epsilon_\gamma^{-\frac{p}{2}},  \hspace{0.95cm}   \,\epsilon^{\rm syn}_{\rm c}<\epsilon_\gamma\,. \cr
}
\end{eqnarray}
}

The synchrotron spectral breaks $\epsilon^{\rm syn}_{\rm m}$ and $\epsilon^{\rm syn}_{\rm c}$, and the maximum flux are given in eq. (\ref{energies_break_syn_off_h}).

\paragraph{SSC light curves}  The SSC light curves for the fast- and slow-cooling regimes, are (respectively): 
{\small
\begin{eqnarray}
\label{FC_syn_bb}
F^{\rm ssc}_{\nu}\propto\cases{
t^{\frac{19}{3}}\, \epsilon_\gamma^{\frac13}\,,\hspace{1.3cm} \epsilon_\gamma<\epsilon^{\rm ssc}_{\rm c},\cr
t^8\,    \,\epsilon_\gamma^{-\frac{p-1}{2}}\,, \hspace{0.8cm}    \epsilon^{\rm ssc}_{\rm c}<\epsilon_\gamma<\epsilon^{\rm ssc}_{\rm m},\,\,\,\,\,\cr
t^{-3p+11}\, \epsilon_\gamma^{-\frac{p}{2}},\, \hspace{0.5cm}  \epsilon^{\rm ssc}_{\rm m}<\epsilon_\gamma\,, \cr
}
\end{eqnarray}
}

{\small
\begin{eqnarray}
\label{FC_syn_bb}
F^{\rm ssc}_{\nu}\propto\cases{
t^9\, \epsilon_\gamma^{\frac13}\,,\hspace{1.8cm} \epsilon^{\rm ssc}_\gamma<\epsilon^{\rm ssc}_{\rm m},\cr
t^{-3p+10}\,    \,\epsilon_\gamma^{-\frac{p-1}{2}}\,, \hspace{0.4cm}    \epsilon^{\rm ssc}_{\rm m}<\epsilon_\gamma<\epsilon^{\rm ssc}_{\rm c},\,\,\,\,\,\cr
t^{-3p+11}\, \epsilon_\gamma^{-\frac{p}{2}},\, \hspace{0.8cm}  \epsilon^{\rm ssc}_{\rm c}<\epsilon_\gamma\,. \cr
}
\end{eqnarray}
}

The SSC spectral breaks $\epsilon^{\rm ssc}_{\rm m}$ and $\epsilon^{\rm ssc}_{\rm c}$, and the maximum flux are given in eq. (\ref{energies_break_ssc_off_h}).

\subsubsection{Homogeneous Medium: Lateral expansion phase}

In the lateral expansion phase ($\Delta \theta\sim 1/\Gamma$),  the fiducial energy becomes  $\tilde{E}=\frac{16\pi}{3}\,m_p(1+z)^{-3}\,n \,\Gamma^{6}\,t^3$ and then the bulk Lorentz factor can be written as:
\be\label{Gamma_j_l}
\Gamma= 3.2\left(\frac{1+z}{1.022}\right)^{\frac{1}{2}}\,n_{-1}^{-\frac{1}{6}}\,\tilde{E}^{\frac{1}{6}}_{50} t_{\rm 100d}^{-\frac{1}{2}}\,. 
\ee

Using eq. (\ref{Gamma_j_l}) in (\ref{FC_syn_bb_g}), (\ref{SC_syn_bb_g}), (\ref{FC_ssc_bb_g}) and (\ref{SC_ssc_bb_g}), we next obtain the synchrotron and SSC light curves of a quasi-spherical outflow in the lateral expansion regime moving through a homogeneous medium.

\paragraph{Synchrotron light curves} The synchrotron light curve for the slow-cooling regime is:
{\small
\begin{eqnarray}
\label{scsyn_t}
F^{\rm syn}_{\nu}\propto\cases{
t^{-\frac{1}{3}} \, \epsilon_\gamma^{\frac13},\hspace{1.4cm} \epsilon_\gamma<\epsilon^{\rm syn}_{\rm m},\cr
t^{-p}\,\epsilon_\gamma^{-\frac{p-1}{2}}, \hspace{1.cm}      \epsilon^{\rm syn}_{\rm m}<\epsilon_\gamma<\epsilon^{\rm syn}_{\rm c},\,\,\,\,\,\cr
t^{-p}\,\epsilon_\gamma^{-\frac{p}{2}}, \hspace{1.3cm}\,\epsilon^{\rm syn}_{\rm c}<\epsilon_\gamma\,. \cr
}
\end{eqnarray}
}

The synchrotron spectral breaks $\epsilon^{\rm syn}_{\rm m}$ and $\epsilon^{\rm syn}_{\rm c}$, and the maximum flux are given in eq. (\ref{energies_break_syn_off_h_l}).

\paragraph{SSC light curves} The SSC light curve for the slow-cooling regime is: 
{\small
\begin{eqnarray}
\label{FC_syn_bb}
F^{\rm ssc}_{\nu}\propto\cases{
t^0\, \epsilon_\gamma^{\frac13}\,,\hspace{1.6cm} \epsilon_\gamma<\epsilon^{\rm ssc}_{\rm m},\cr
t^{\frac{1-3p}{2}}\,    \,\epsilon_\gamma^{-\frac{p-1}{2}}\,, \hspace{0.4cm}    \epsilon^{\rm ssc}_{\rm m}<\epsilon_\gamma<\epsilon^{\rm ssc}_{\rm c},\,\,\,\,\,\cr
t^{\frac{2-3p}{2}}\, \epsilon_\gamma^{-\frac{p}{2}},\, \hspace{0.8cm}  \epsilon^{\rm ssc}_{\rm c}<\epsilon_\gamma\,. \cr
}
\end{eqnarray}
}

The SSC spectral breaks $\epsilon^{\rm ssc}_{\rm m}$ and $\epsilon^{\rm ssc}_{\rm c}$, and the maximum flux are given in eq.  (\ref{energies_break_ssc_off_h_l}).

\subsubsection{Wind-like Medium}
Considering the Blandford-McKee solution for a wind-like medium,  the bulk Lorentz factor can be written as:
{\small
\be\label{Gamma_o_w}
\Gamma=2.4\times 10^3\,\left(\frac{1+z}{1.022}\right)^{\frac{1}{2}} \xi^{-1} \,A^{-\frac{1}{2}}_{\star,-1}\,\tilde{E}^{\frac{1}{2}}_{50}\,\theta_{j,5^\circ}^{-1}\,\Delta\theta_{15^\circ}\,t^{-\frac{1}{2}}_{\rm 1d}\,,
\ee
}
where $\tilde{E}$ is the same as that for the quasi-spherical outflow in a wind-like medium.

\paragraph{Synchrotron light curves} The synchrotron light curves for the fast- and the slow-cooling regimes are (respectively): 
{\small
\begin{eqnarray}
\label{scsyn_t}
F^{\rm syn}_{\nu}\propto\cases{
t^{\frac13} \, \epsilon_\gamma^{\frac13},\hspace{1.5cm} \epsilon_\gamma<\epsilon^{\rm syn}_{\rm c},\cr
t^2\,\epsilon_\gamma^{-\frac{p-1}{2}}, \hspace{1.0cm}     \epsilon^{\rm syn}_{\rm m}<\epsilon_\gamma<\epsilon^{\rm syn}_{\rm c},\,\,\,\,\,\cr
t^{3-p}\,\epsilon_\gamma^{-\frac{p}{2}},\hspace{1.0cm} \,\epsilon^{\rm syn}_{\rm c}<\epsilon_\gamma\,, \cr
}
\end{eqnarray}
}

{\small
\begin{eqnarray}
\label{scsyn_t}
F^{\rm syn}_{\nu}\propto\cases{
t^{\frac53} \, \epsilon_\gamma^{\frac13},\hspace{1.95cm} \epsilon_\gamma<\epsilon^{\rm syn}_{\rm m},\cr
t^{-p+2}\,\epsilon_\gamma^{-\frac{p-1}{2}}, \hspace{1.0cm}\epsilon^{\rm syn}_{\rm m}<\epsilon_\gamma<\epsilon^{\rm syn}_{\rm c},\,\,\,\,\,\cr
t^{3-p}\,\epsilon_\gamma^{-\frac{p}{2}},\hspace{1.4cm}\,\epsilon^{\rm syn}_{\rm c}<\epsilon_\gamma\,. \cr
}
\end{eqnarray}
}

The synchrotron spectral breaks $\epsilon^{\rm syn}_{\rm m}$ and $\epsilon^{\rm syn}_{\rm c}$, and the maximum flux are given in eq. (\ref{energies_break_syn_off_w}).

\paragraph{SSC light curves} The SSC light curves for the fast- and the slow-cooling regimes are (respectively): 
{\small
\begin{eqnarray}
\label{FC_syn_bb}
F^{\rm ssc}_{\nu,off}\propto\cases{
t^{-\frac{5}{3}}\, \epsilon_\gamma^{\frac13}\,,\hspace{1.2cm} \epsilon_\gamma<\epsilon^{\rm ssc}_{\rm c},\cr
t^{\frac52}\,    \,\epsilon_\gamma^{-\frac{p-1}{2}}\,, \hspace{0.8cm}    \epsilon^{\rm ssc}_{\rm c}<\epsilon_\gamma<\epsilon^{\rm ssc}_{\rm m},\,\,\,\,\,\cr
t^{-\frac{3p}{2}+4}\, \epsilon_\gamma^{-\frac{p}{2}},\, \hspace{0.5cm}  \epsilon^{\rm ssc}_{\rm m}<\epsilon_\gamma\,, \cr
}
\end{eqnarray}
}

{\small
\begin{eqnarray}
\label{FC_syn_bb}
F^{\rm ssc}_{\nu}\propto\cases{
t\, \epsilon_\gamma^{\frac13}\,,\hspace{2.05cm} \epsilon_\gamma<\epsilon^{\rm ssc}_{\rm m},\cr
t^{-\frac32 (p-1)}\,    \,\epsilon_\gamma^{-\frac{p-1}{2}}\,, \hspace{0.4cm}    \epsilon^{\rm ssc}_{\rm m}<\epsilon_\gamma<\epsilon^{\rm ssc}_{\rm c},\,\,\,\,\,\cr
t^{-\frac{3p}{2}+4}\, \epsilon_\gamma^{-\frac{p}{2}},\, \hspace{0.95cm}  \epsilon^{\rm ssc}_{\rm c}<\epsilon_\gamma\,. \cr
}
\end{eqnarray}
}

The synchrotron spectral breaks $\epsilon^{\rm ssc}_{\rm m}$ and $\epsilon^{\rm ssc}_{\rm c}$, and the maximum flux are given in eq. (\ref{energies_break_ssc_off_w}).

\subsection{Multiwavelength Light Curves}

\begin{figure}
{ \centering
\resizebox*{1\textwidth}{0.65\textheight}
{\includegraphics{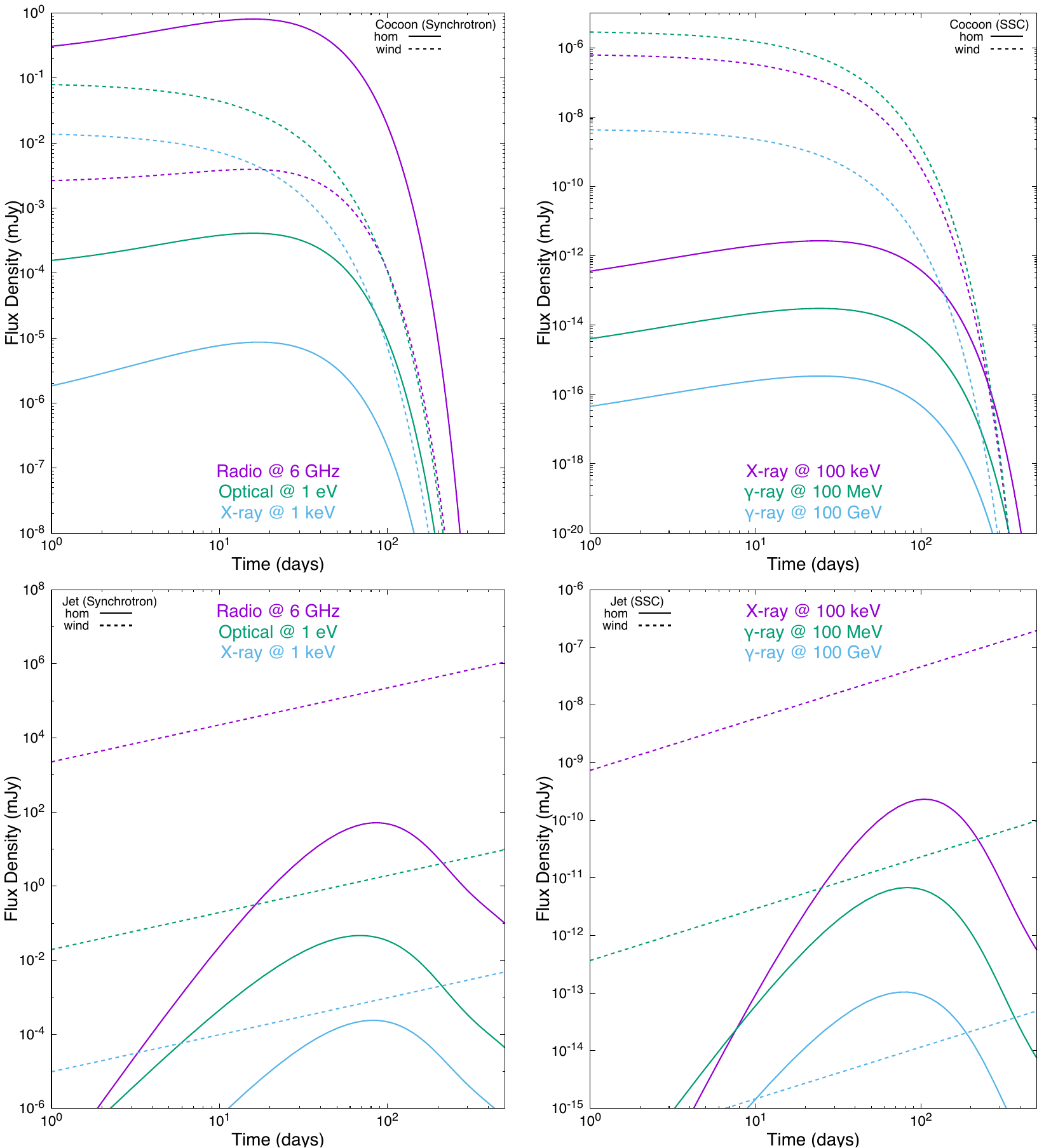}}
}
\caption{Synchrotron (left-hand panels) and SSC (right-hand panels) light curves produced by a quasi-spherical outflow (upper panels) or an off-axis jet (lower panels). The purple, green and blue lines correspond to 6 GHz, 1 eV and 1 keV in the left-hand panels and 100 keV, 100 MeV and 100 GeV in right-hand panels. The continuous lines correspond to a homogeneous medium, and the dashed lines to a wind-like medium. The values used are $E=5\times 10^{49}\,{\rm erg}$, $n=5\times 10^{-4}\,{\rm cm^{-3}}$, $A_{\star}=10^{-4}$, $\varepsilon_B=5\times10^{-4}$, $\varepsilon_e=0.1$, $\Delta \theta=18^\circ$, $\theta_j=7^\circ$, $\alpha=3.0$, p=2.25 and D=100 Mpc.}
\label{fig2:cocoon_ism}
\end{figure}

\begin{figure}
{ \centering
\resizebox*{\textwidth}{0.35\textheight}
    {\includegraphics{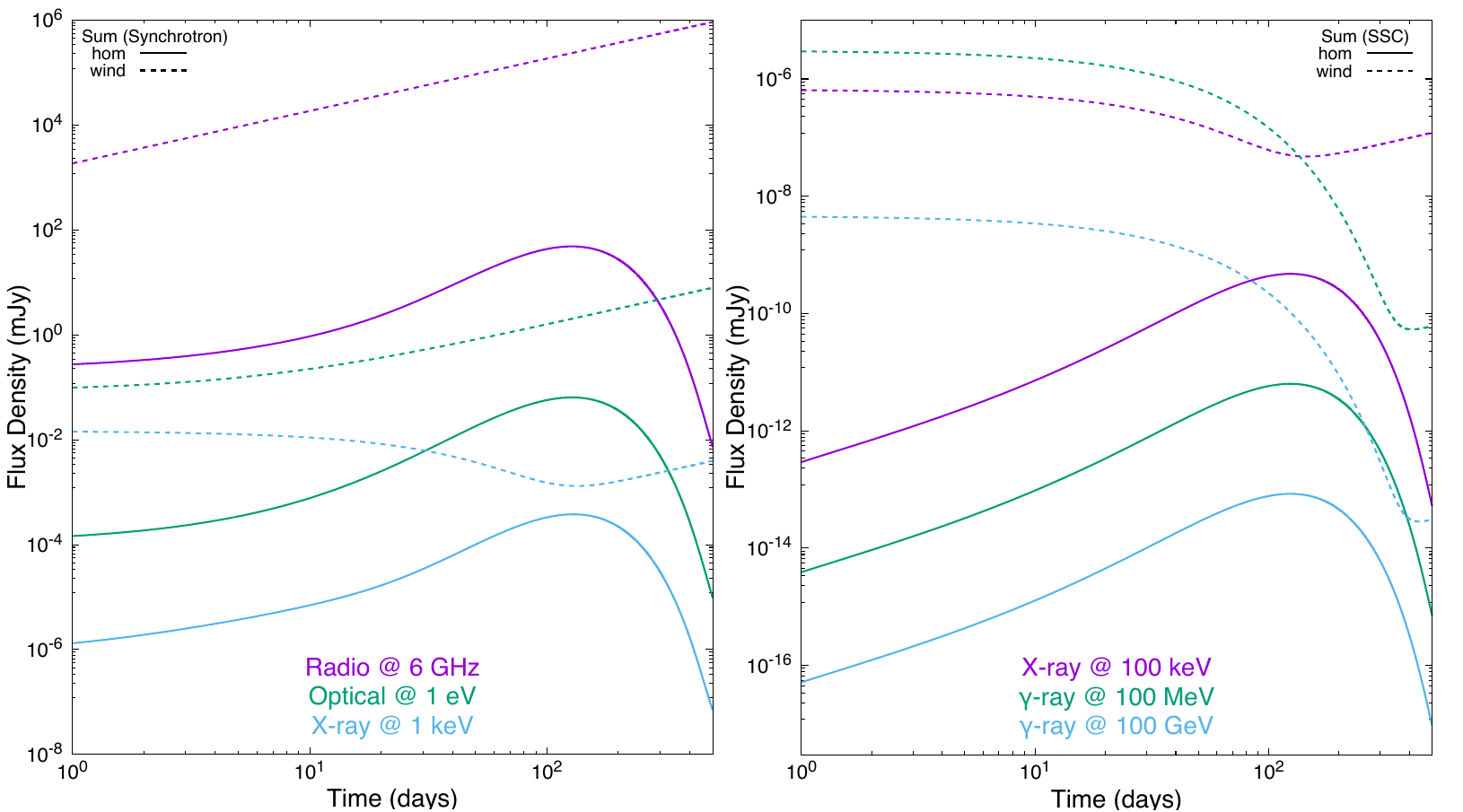}}
}
\caption{Synchrotron (left-hand panels) and SSC (right-hand panels) light curves produced by a quasi-spherical outflow and an off-axis jet. The colors, continuous or dashed lines and values used, are the same as those in Fig.\ref{fig2:cocoon_ism}}
\label{fig4:jet+cocoon}
\end{figure}

Figures \ref{fig2:cocoon_ism} and  \ref{fig4:jet+cocoon} show the resulting $\gamma$-ray, X-ray, optical and radio light curves of  the synchrotron and SSC forward-shock radiation produced by a quasi-spherical outflow, or an off-axis jet, and decelerating either in a homogeneous or a wind-like medium. The purple, green and blue lines correspond to 6 GHz, 1 eV and 1 keV in the left-hand panels and 100 keV, 100 MeV and 100 GeV in the right-hand panels, respectively, and the standard values of GRB afterglows were used.\footnote{$E=5\times 10^{49}\,{\rm erg}$, $n=5\times 10^{-4}\,{\rm cm^{-3}}$, $A_{\star}=10^{-4}$, $\varepsilon_B=5\times 10^{-4}$,  $\varepsilon_e=0.1$, $\Delta \theta=18^\circ$, $\theta_j=7^\circ$,  $\alpha_s=3.0$, p=2.25 and D=100 Mpc.} The figures show the light curves from 1 to 1000 days for six electromagnetic bands: $\gamma$-ray at 100 GeV, 100 MeV and 100 keV, X-ray at 1 keV, optical at 1 eV  and radio at 6 GHz. It is worth noting that the synchrotron and SSC light curves shown in the previous figures lie in the slow cooling regime.\\

For density values ($n=10^{-4}\,{\rm cm^{-3}}$ for the homogeneous medium, and $A_\star=10^{-4}$ for the wind-like medium), the synchrotron and SSC fluxes produced by the quasi-spherical outflow are noticeably larger in a wind-like medium than in a homogeneous medium. The synchrotron light curves in a wind-like medium are 2 - 7 orders of magnitude larger than in a homogeneous medium (see the upper-left panel in Figure \ref{fig2:cocoon_ism}). Meanwhile, the SSC light curves in a wind-like medium are 6-10 orders of magnitude larger than in a homogeneous medium (see the upper-right panel in Figure \ref{fig2:cocoon_ism}). The disparity between the synchrotron and SSC emission depends on the energy band, the timescale considered, and the chosen $A_\star$ parameter (as $A_\star$ increases, so do the fluxes in the wind-like medium). We also find that the synchrotron and SSC light curves of a quasi-spherical outflow expanding through a homogeneous or a wind-like medium behave differently. The synchrotron and SSC fluxes of a quasi-spherical outflow in a homogeneous medium increase gradually during the first $\sim$ 20-50 days, then reach a maximum value and decrease afterwards. In a wind-like medium, on the other hand, the quasi-spherical outflow emission decreases monotonically in all electromagnetic bands (except in the radio band where it increases). For the off-axis jet we find that the synchrotron and SSC fluxes produced have similar values (see the bottom panels of Figure \ref{fig2:cocoon_ism}). The synchrotron and SSC fluxes of an off-axis jet in a homogeneous medium increase during the first $\sim$ 100 days, then reach a maximum value and decrease rapidly afterwards. In a wind-like medium, the jet's emission increases monotonically in all electromagnetic bands. Comparing the synchrotron emission of the quasi-spherical outflow with that of the off-axis jet (in the same ambient medium regime), it is clear that the emission of the quasi-spherical outflow is stronger than that of the off-axis jet during the first $\sim$ 10 - 20 days, and then weaker during the next $\sim$ 80 days. A similar behavior occurs for the SSC. In the stellar-wind medium, except for the radio band, the flux generated by the quasi-spherical outflow governs during the timescale considered. In the case of the radio band, during the first $\sim$ 80 days the radio flux emitted by the quasi-spherical outflow prevails, afterwards the radio flux from the off-axis jet dominates.\\ 

Figure \ref{fig4:jet+cocoon} shows the total light curves (built by the sum of the emission from the quasi-spherical outflow and the jet) for the synchrotron and SSC cases produced in a homogeneous or a wind-like medium. The light curves produced in a homogeneous medium increase during the first $\sim$100 days, then reach their respective maximum, and decrease afterwards. The synchrotron light curves for the wind-like medium have different behaviours; whereas the radio flux is an increasing function, the optical and X-ray fluxes are decreasing functions. Meanwhile, regardless of the energy band, the SSC light curves from the wind-like medium, are decreasing functions. It is worth noting that the SSC emission produced in wind-like medium is at least four orders of magnitude larger than that produced in a homogeneous medium.

The standard synchrotron forward-shock model predicts that the spectral evolution of frequencies evolves as $\epsilon_{\rm m} \propto t^{-\frac32}$ and $\epsilon_{\rm m} \propto t^{\frac12}$ in wind-like medium and  $\epsilon_{\rm m} \propto t^{-\frac32}$ and $\epsilon_{\rm m} \propto t^{-\frac12}$ in a homogeneous medium. \citet{1999ApJ...524L..47G} found a new component during the prompt phase, different to the Band function, for GRB 980923. The analysis revealed that the spectral evolution of this component was similar to that described by the evolution of the cooling frequency in the synchrotron forward-shock model $\epsilon_{\rm c}\propto t^{-0.5}$, thus arguing that external shocks can be created during the prompt phase. Subsequently, several papers \citep{2006MNRAS.369..311Y, 2018arXiv180207328V, 2019ApJ...871..200F, 2005ApJ...635L.133B} were written in this direction in order to identify the early afterglow phase during the gamma-ray prompt emission. In this manuscript, we provide useful tools to identify this early afterglow in a homogeneous or wind-like medium (see appendix A). For instance, the evolution of the spectral component generated by the deceleration of the quasi-spherical outflow in a wind-like medium is $\epsilon_{\rm m}\propto t^{-\frac{3}{\alpha_s+8}}$ and $\epsilon_{\rm c}\propto t^{-\frac{1-\alpha_s}{\alpha_s+8}}$, and in a homogeneous medium it is $\epsilon_{\rm m}\propto t^{-\frac{3}{\alpha_s+8}}$ and $\epsilon_{\rm c}\propto t^{-\frac{1-\alpha_s}{\alpha_s+8}}$. In the particular case of $\alpha_{\rm s}=0$, the temporal evolution of the synchrotron spectral breaks derived in \cite{1998ApJ...497L..17S, 2016ApJ...831...22F,  1999ApJ...519L.155D, 2003MNRAS.341..263H, 2000ApJ...537..785D, 2002ApJ...570L..61G, 1999A&AS..138..491R} are recovered. The evolution of the synchrotron spectral breaks generated by the deceleration of the off-axis jet in a wind-like medium is  $\epsilon_{\rm m}\propto t^{\frac{3}{\alpha_s+8}}$ and  $\epsilon_{\rm c}\propto t^{\frac{\alpha_s+3}{\alpha_s+4}}$, and in a homogeneous medium it is $\epsilon_{\rm m}\propto t^{-2}$ and  $\epsilon_{\rm c}\propto t^2$.  We emphasize that we ignore  the scattering from the jet when we calculate the SSC from the quasi-spherical outflow.

\section{Application: GRB 170817A}

\begin{figure}
	{ \centering
		\resizebox*{\textwidth}{0.7\textheight}
		{\includegraphics[angle=-90]{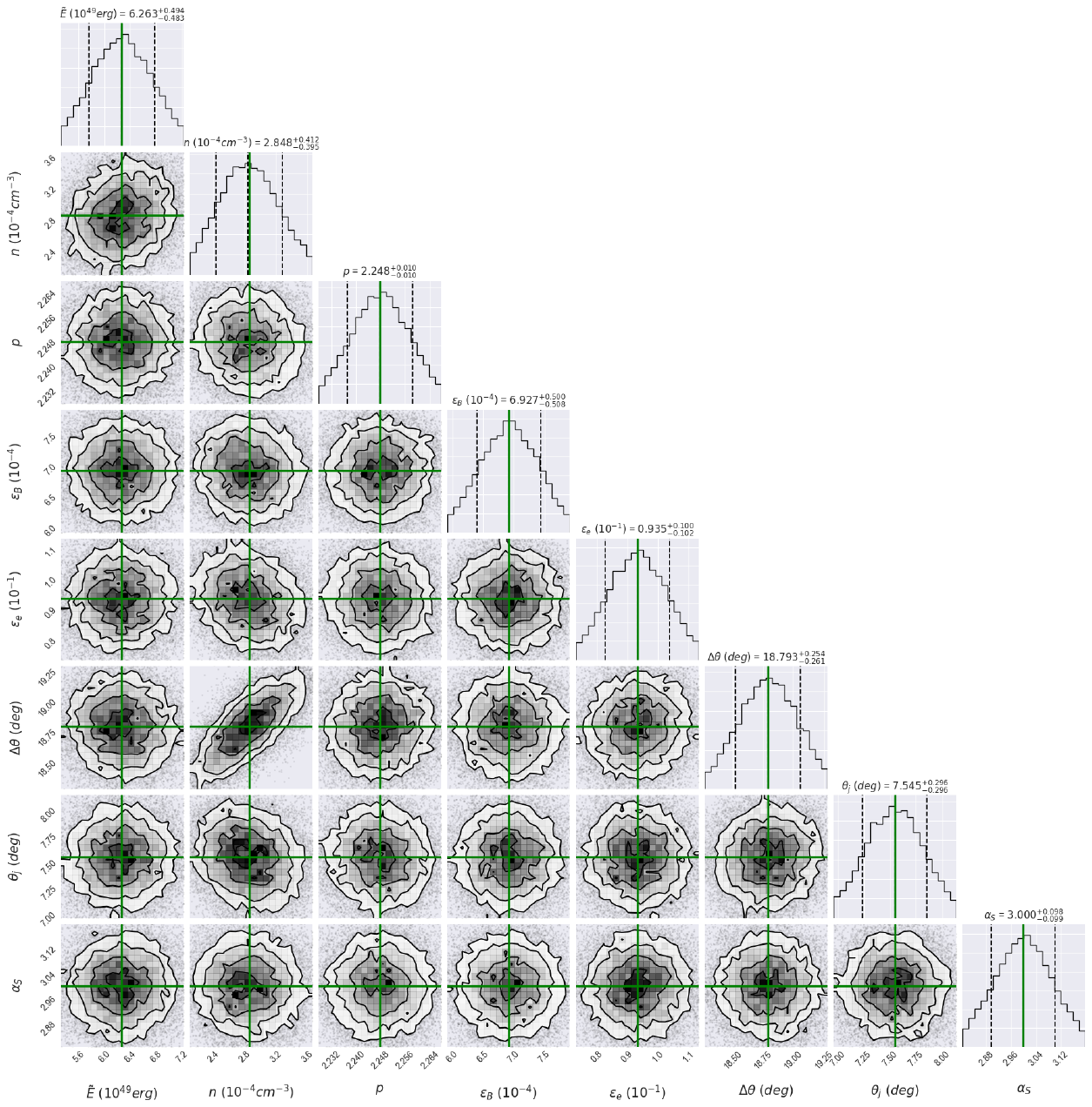}}
	}
	\caption{Best fit results for the light curves at 3 GHz using the proposed model and the MCMC calculations for GRB 170817A. The ``corner plots" exhibit  the results obtained from the MCMC simulation. Labels above the 1-D KDE plot illustrate the 15\%, 50\% and 85\% quantiles for all parameters. The best-fit values are shown in green and reported in Table \ref{table1:parameters}.}
	\label{fig3:param_late_3R}
\end{figure}

To find the best-fit values that describe the non-thermal emission of GRB 170817A,  we use a Markov-Chain Monte Carlo (MCMC) code \cite[see][]{2019ApJ...871..200F}. The MCMC code calculates the synchrotron emission of a quasi-spherical outflow and an off-axis jet and is described by a set of eight parameters, \{$\tilde{E}$, n, p, $\theta_j$, $\Delta\theta$, $\varepsilon_B$, $\varepsilon_e$, $\alpha_s$\}. A total of 17600 samples with 5150 tuning steps were run. The best fit parameters of $\Delta\theta$, $p$, ${\rm n}$, $\varepsilon_{\rm B}$, $\varepsilon_e$,  $\alpha_s$ and $\tilde{E}$ are displayed in Figures \ref{fig3:param_late_3R} (radio: 3 GHz, 6 GHz; optical: 1 eV; and X-ray: 1 keV). The best-fit values for GRB 170817A are reported in Table \ref{table1:parameters}. The obtained values are consistent with those reported by other authors \citep{2017arXiv171111573M, 2019ApJ...871..200F,2018ApJ...867...95H, 2017Sci...358.1559K, 2017MNRAS.472.4953L, 2017arXiv171203237L}.  We note that the synchrotron flux equations are degenerate in these parameters such that for a completely different set of parameters the same results can be obtained. Therefore, our result is not unique, but is only one possible solution to GRB 170817A.\\

\begin{figure}
{ \centering
\resizebox*{0.5\textwidth}{0.35\textheight}
{\includegraphics{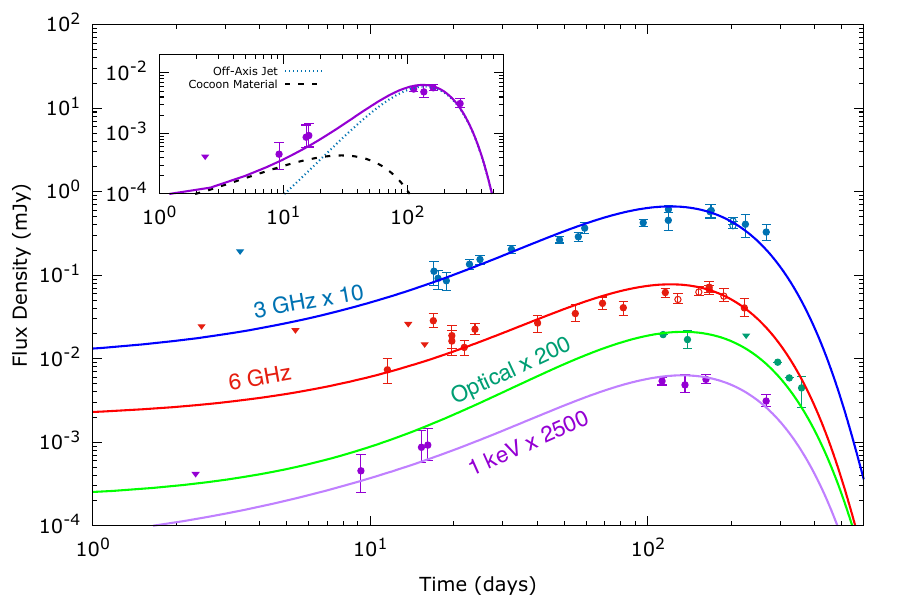}}
\resizebox*{0.5\textwidth}{0.35\textheight}
{\includegraphics{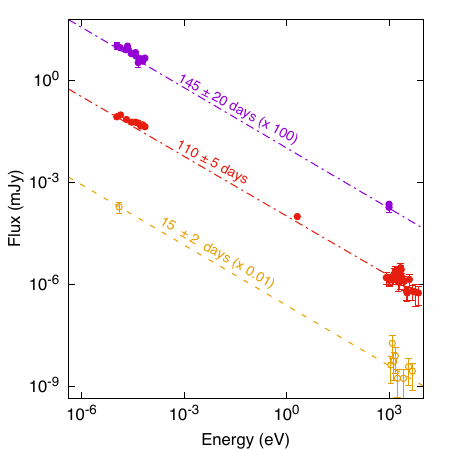}}
}
\caption{Left: The best-fit light curves obtained using the synchrotron emission from a quasi-spherical outflow and an off-axis jet decelerated in a homogeneous medium. These light curves are exhibited at different energy bands with their respective observations (points). The radio energy band at 3 GHz is shown in cyan, the radio energy band at 6 GHz is shown in red, the optical band at 1 eV is shown in green, and the X-ray at 1 keV in purple. A zoom of the X-ray light curve and the emission produced by the quasi-spherical outflow and off-axis jet is also shown (upper-left). The data points are the observations, see the text for their references. Right: The best-fit SEDs of the X-ray (red), optical (green), and radio (blue) afterglow observations at 15 $\pm$ 2, 110 $\pm$ 5, and 145 $\pm$ 20 days respectively. The values which best describe the light curves and the SED are reported in Table \ref{table1:parameters}.}
\label{fig4:afterglow}
\end{figure}

Figure \ref{fig4:afterglow} shows the obtained light curves (left panel) and the spectral energy distributions (SED, right panel) of the X-ray, optical and radio bands of GRB 170817A, the data points were taken from \citet{2017Natur.547..425T, 2017ApJ...848L..20M, 2017ATel11037....1M, 2018ATel11242....1H, 2018arXiv180106516T,2018arXiv180103531M,  2017arXiv171005435H, 2017arXiv171111573M, 2017ApJ...848L..21A}. The light curves are shown in radio wavelengths at 3 and 6 GHz, optical band at 1 eV and X-rays at 1 keV. The SEDs are exhibited at $15\pm 2$, $110\pm 5$ and $145 \pm 20$ days. 

The multiwavelength data (radio wavelengths at 3 and 6 GHz, optical band at 1 eV and X-rays at 1 keV) were described through the best-fit curves of synchrotron radiation emitted from the deceleration of the quasi-spherical outflow and the off-axis jet. The maximum value of the flux density in each band is interpreted by the broadening of the beaming cone of the radiation. It occurs when the off-axis jet has slowed down and  expanded laterally. A zoom of the X-ray light curve, with the correspondent emission produced by the quasi-spherical outflow and the off-axis jet, is also shown in the left-hand panel. The dashed-black line shows the contribution of the quasi-spherical outflow and the dotted-blue line shows the contribution of the off-axis jet. This figure shows that emission from the quasi-spherical outflow dominates during the $\sim$ 20 days and the emission from the off-axis jet dominates after the $\sim$ 60 days.\\

Using the values of the best-fit parameters reported in Table \ref{table1:parameters} and eq. (\ref{Gamma_c}), we find that  the bulk Lorentz factor is $\Gamma_{\rm c}\simeq 3.1 \left(t/15\,d\right)^{-0.24}$ and the equivalent kinetic energy is $E_{\rm obs, k}\simeq 3.31\,\times 10^{47}\,{\rm erg}$. Using the previous values, we obtain that the efficiency to convert the kinetic energy to gamma-ray energy is $\sim 16\%$. This value is consistent with the range of values reported in afterglows \cite[e.g. see,][]{2004IJMPA..19.2385Z, 2015PhR...561....1K}. The cooling and characteristic spectral breaks are $\epsilon_{\rm c}\sim 22.3$ eV and $\epsilon_{\rm m}\sim 1.1\times10^{-2}$ GHz, respectively, at 15 days. This result is consistent with the evolution of synchrotron radiation in the slow-cooling regime of the quasi-spherical outflow in a homogeneous medium, where the X-ray, optical and radio fluxes are described by the third and second power-law segment in eq. (\ref{SC_syn_bb_c}). The X-ray, optical and radio fluxes increase as $F_\nu\propto t^{0.15}$, peak at $\sim$ 20 days, and then evolve as $F_\nu\propto t^{-0.76}$ and  $\propto t^{-1.03}$. On the other hand, the optical and radio fluxes continue evolving as $F_\nu\propto t^{-0.76}$. Given the values of the best-fit parameters reported in Table  \ref{table1} and  eq. (\ref{Gamma_o}), we find that the bulk Lorentz factor of the relativistic jet reaches $\Gamma_{\rm j}\simeq 5.3 \left(t/100\,d\right)^{-\frac32}$. The cooling spectral break $\epsilon_{\rm c}\sim2.6$ keV is above the X-ray band, and its characteristic break $\epsilon_{\rm m}=0.04$ GHz is below the radio band at 100 days. As for the quasi-spherical outflow, this result is consistent with the evolution of synchrotron radiation in the slow-cooling regime of an off-axis jet expanding in a homogeneous medium (where the X-ray, optical and radio fluxes are described by the second power-law segment in eq. \ref{scsyn_t}). During this period, the observed flux increases as $F_\nu\propto t^{4.2}$ as predicted in \cite{2018arXiv180109712N}. The X-ray, optical, and radio fluxes peak at $\sim$ 140 days, and then evolve as $F_\nu\propto t^{-2.2}$. It is worth noting that for a time scale of seconds, an equivalent kinetic energy above $\sim 5\times 10^{52}\,{\rm erg}$, a circumburst density higher than $1\,{\rm cm^{-3}}$, and equipartition parameters $\varepsilon_{\rm B}\sim 0.1$, $\varepsilon_{\rm e}\sim 0.1$, the synchrotron and SSC light curves would lie in the fast-cooling regime.\\ 

\begin{figure}[h!]
{ \centering
	\resizebox*{\textwidth}{0.4\textheight}
      {\includegraphics{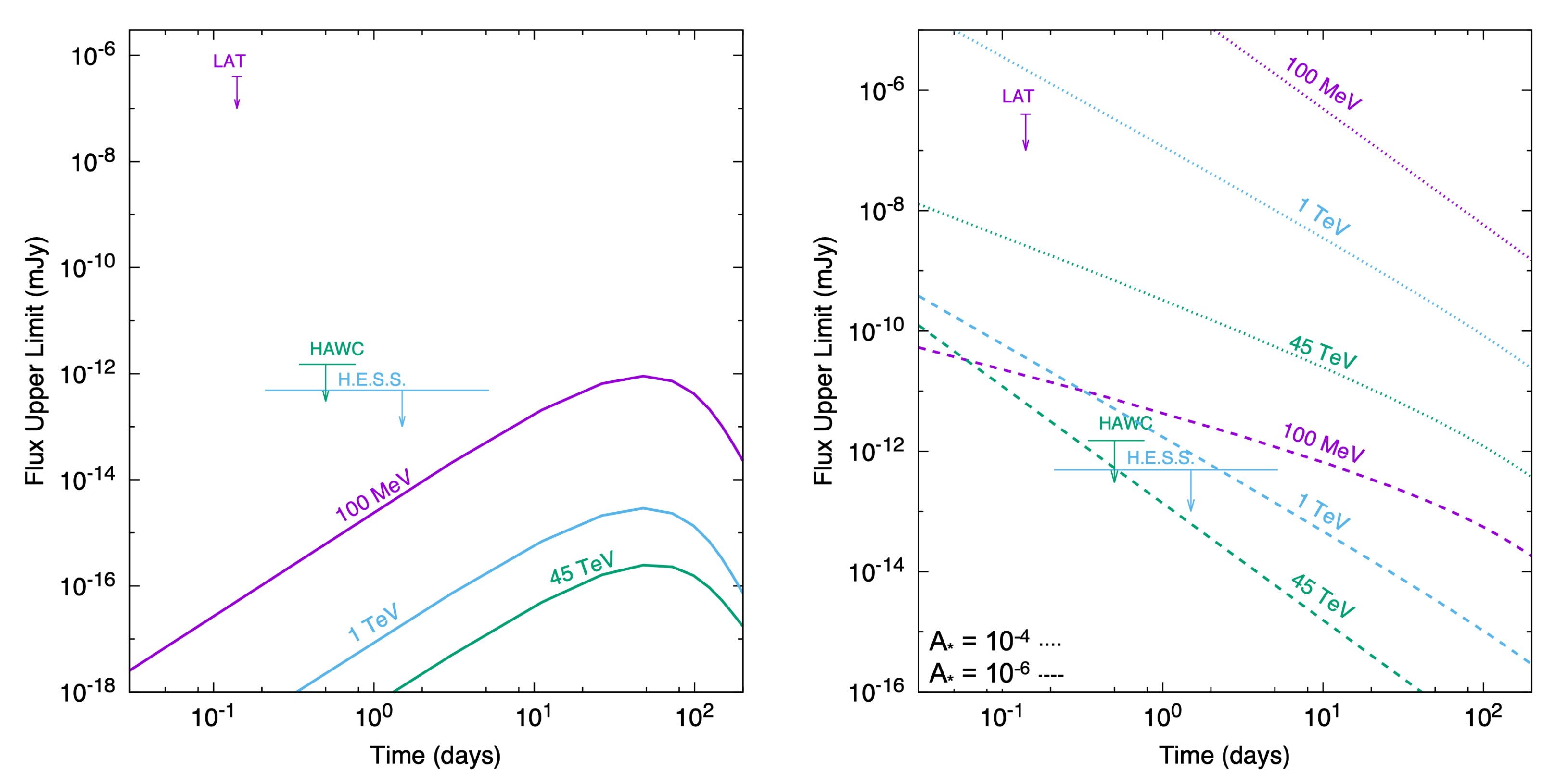}}
}
\caption{Upper limits derived with the Fermi-LAT, HAWC and H.E.S.S. observatories with the SSC model from an off-axis jet and quasi-spherical outflow. In the left-hand panel we have used the values found to describe the X-ray, optical, and radio light curves of GRB 170817A moving in a homogeneous medium \cite[see Table 5 in][]{2019ApJ...871..123F}. In the right-hand panel we have assumed that the off-axis jet and quasi-spherical outflow evolve in a wind-like medium.}
\label{LC_GRB170817A_HE}
\end{figure}

The results reported in the radio energy band by \citet{2018Natur.561..355M} reported superluminal motion, with an apparent speed of $\sim$ 4 at almost 150 days (between  75 and 230 days after the GBM trigger), which implies that a relativistic jet is present. This result was confirmed by the radio observations performed 207.4 days after the NS fusion \citep{2018arXiv180800469G}. These observations provide compelling evidence that the progenitor of the GW170817 event ejected a structured relativistic jet with  a bulk Lorentz factor of $\sim$ 4 (at the time of measurement), observed from a viewing angle of $20^\circ\pm5^\circ$. The model proposed in this manuscript  is consistent with the results obtained in the radio wavelengths, which at earlier times show that the non-thermal emission is dominated by the slower quasi-spherical outflow material, and at later times, the non-thermal emission ($\gtrsim$ 80 days post-merger) is dominated by a relativistic off-axis jet. Considering the values of $\Delta \theta\simeq18^\circ$ and $\theta_j\simeq 7^\circ$ reported in Table \ref{table1:parameters}, the value of the viewing angle $\theta_{\rm obs} \sim 25^\circ$ is found, which agrees with that reported in \cite{2018Natur.561..355M}.\\   
Using values obtained with the MCMC simulation for GRB 170817A, we calculate the correspondent fluxes of the SSC model to compare them with Fermi-LAT, HAWC, and H.E.S.S. upper limits. The left-hand panel from Figure \ref{LC_GRB170817A_HE} shows the obtained SSC light curves (solid lines) as well as the upper limits obtained by  Fermi-LAT, HAWC, and H.E.S.S..The light curves at 100 MeV (purple), 1 TeV (blue) and 45 TeV (green), were obtained using the values reported in Table \ref{table1:parameters}. The effect of the extragalactic background light absorption model of \citet{2017A&A...603A..34F} was used. The obtained SSC flux at different energy bands agrees with the LAT, H.E.S.S.  and  HAWC observatories. The right-hand panel of Figure \ref{LC_GRB170817A_HE} shows the SSC light curves in a wind-like medium. If the SSC flux would have been emitted in a wind-like medium, it could have been observed by LAT, H.E.S.S. or HAWC Observatories. For instance, with $A_\star=10^{-4}$ the SSC electromagnetic signal would have been detected in these observatories, but not with $A_\star=10^{-6}$. This result is very interesting since the material that surrounds the progenitor of the short GRB may be affected by the wind and launched material produced during the merger of the NSs  \citep[e.g. see,][]{2018ApJ...863L..34B}. 

\section{Conclusions}

We have derived an analytic model of the forward-shock, produced by the ejection of material (after the merger of two NSs), and which is moving either in a homogeneous or a wind-like medium. Explicitly, we have obtained the SSC and synchrotron light curves in the fast- and slow-cooling regimes during the relativistic and lateral expansion phases in the fully adiabatic regime with arbitrary line of sights for an observer. We focus our model in the emission from a quasi-spherical outflow that is viewed on-axis and an off-axis relativistic (top-hat) jet, and we describe the extended X-ray, optical and radio emission exhibited in GRB170817A. We find that the SSC and synchrotron light curves produced by a quasi-spherical outflow can be expressed when the equivalent kinetic energy is  $\tilde{E}\, \Gamma^{-\alpha_s}$, and the light curves produced by an off-axis jet when the equivalent kinetic energy is $2\tilde{E}/\theta^2_j$. In the particular case of $\alpha_{\rm s}=0$, the SSC and synchrotron light curves derived in \cite{1998ApJ...497L..17S, 2016ApJ...831...22F,  1999ApJ...519L.155D, 2003MNRAS.341..263H, 2000ApJ...537..785D, 2002ApJ...570L..61G, 1999A&AS..138..491R} are recovered. The flux of a quasi-spherical outflow which is expanding in a wind-like medium is several orders of magnitude larger than that generated in a homogeneous medium. The latter is also the case for the off-axis jet at early times ($t\lesssim 15$ days).  The flux produced by the quasi-spherical outflow peaks before the flux of the off-axis jet and dominates during the first 10-20 days (compared to that from the off-axis jet). At later times ($\gtrsim 100$ days), the emission of the off-axis jet peaks and dominates. We show that the evolution of the spectral component generated by the deceleration of the quasi-spherical outflow in a wind-like medium is $\epsilon_{\rm m}\propto t^{-\frac{3}{\alpha_s+8}}$ and $\epsilon_{\rm c}\propto t^{-\frac{1-\alpha_s}{\alpha_s+8}}$,  and in a homogeneous medium is $\epsilon_{\rm m}\propto t^{-\frac{3}{\alpha_s+8}}$ and $\epsilon_{\rm c}\propto t^{-\frac{1-\alpha_s}{\alpha_s+8}}$. The evolution of the synchrotron spectral breaks generated by the deceleration of the off-axis jet in a wind-like medium is  $\epsilon_{\rm m}\propto t^{\frac{3}{\alpha_s+8}}$ and  $\epsilon_{\rm c}\propto t^{\frac{\alpha_s+3}{\alpha_s+4}}$, and in a homogeneous medium is  $\epsilon_{\rm m}\propto t^{-2}$ and  $\epsilon_{\rm c}\propto t^2$.
 
In order to interpret the non-thermal emission detected from GRB 170817A, we calculated the synchrotron and SSC contributions from both the off-axis jet and the quasi-spherical outflow moving through a homogeneous medium using a MCMC code. We ran a large set of samples to find the best-fit values of $\tilde{E}$, n, p, $\theta_j$, $\Delta \theta$, $\varepsilon_B$, $\varepsilon_e$, $\alpha_s$ that describe the non-thermal emission.  Our model is consistent with the results obtained in the radio wavelengths. We find that at earlier times the non-thermal emission is dominated by the slower quasi-spherical outflow, and at later times, the non-thermal emission ($\gtrsim$ 80 days post-merger) is mainly produced by a relativistic off-axis jet. For the quasi-spherical outflow, we found that the bulk Lorentz factor is mildly relativistic which corresponds to an equivalent kinetic efficiency of $\sim 16\%$. The cooling spectral breaks found for the cocoon and off-axis jet are consistent with synchrotron radiation in the slow-cooling regime. During the first $\sim$ 120 days, we find that the observed flux generated by the deceleration of the off-axis jet increases as $F_\nu\propto t^{\alpha}$ with $\alpha>3$. Using the values obtained with the MCMC simulation for GRB 170817A, we found that the SSC light curves are consistent with the upper limits placed by Fermi-LAT, HAWC and  H.E.S.S.. For a wind-like medium we found that an electromagnetic signature would have been detected by these high-energy observatories.


%


\newpage
\begin{table}
\centering \renewcommand{\arraystretch}{2}\addtolength{\tabcolsep}{3pt}\label{table1}
\caption{Best-fit values for GRB 170817A}
\label{table1:parameters}
\begin{tabular}{ l  l  l  l l c }
\hline
\hline
{\large   Parameters}	& 	{\large  Median}  \\ 

\hline \hline
\\
\small{$\tilde{E}\, (10^{49}\,{\rm erg})$}	\hspace{1cm}&   \small{$6.263^{+0.494}_{-0.485}$} \\
\small{${\rm n}\,\, (10^{-4}\,{\rm cm^{-3}}$ ) }\hspace{1cm}	&  \small{$2.848^{+0.412}_{-0.395}$} \\
\small{${\rm p}$}	\hspace{1cm}&  \small{$2.248^{+0.010}_{-0.010}$}  \hspace{0.7cm} \\
\small{$\theta_j$\,({\rm deg})}	\hspace{1cm}&  \small{$7.545^{+0.296}_{-0.296}$}   \\
\small{$\Delta \theta$\,({\rm deg})}	\hspace{1cm}&  \small{$18.793^{+0.254}_{-0.261}$}	 \hspace{0.7cm} \\
\small{$\varepsilon_B\,\,(10^{-4})$}   \hspace{1cm}&\small{$6.927^{+0.500}_{-0.508}$} \\
\small{$\varepsilon_{e}\,\,(10^{-1})$}	 \hspace{1cm}&\small{$0.935^{+0.100}_{-0.102}$}    \\
\small{${\alpha_s}$}  \hspace{1cm}&\small{$3.000^{+0.098}_{-0.099}$} \\
\hline
\end{tabular}
\end{table}

\clearpage

\clearpage

\newpage
\appendix
\section{A.  Quasi-spherical outflow}

\subsection{Homogeneous medium}
Using the bulk Lorentz factor (eq. \ref{Gamma_c}),  we derive and show the observable quantities when the quasi-spherical outflow is decelerated in a homogeneous medium.

\paragraph{Synchrotron radiation.} In this case, the minimum and cooling Lorentz factors are given by
{\small
\begin{eqnarray}\label{elec_Lorentz_c_h}
\gamma_{\rm m}&\simeq&  1.1\times 10^{3}\, \left(\frac{1+z}{1.022}\right)^{\frac{3}{\alpha_s+8}}\,g(p)\,\varepsilon_{e,-1}\,\,n_{-1}^{-\frac{1}{\alpha_s+8}}\,\tilde{E}^{\frac{1}{\alpha_s+8}}_{50} t_{\rm 1d}^{-\frac{3}{\alpha_s+8}}\cr
\gamma_{\rm c}&\simeq&  6.3\times 10^{5}\,{\rm GHz}\,\, \left(\frac{1+z}{1.022}\right)^{\frac{\alpha_s-1}{\alpha_s+8}}\,(1+Y)^{-1} \varepsilon^{-1}_{B,-1}\,\,n_{-1}^{-\frac{\alpha_s+5}{\alpha_s+8}}\,\tilde{E}^{-\frac{3}{\alpha_s+8}}_{50} t_{\rm 1d}^{\frac{1-\alpha_s}{\alpha_s+8}}\,.
\end{eqnarray}
}

Given the synchrotron radiation ($\epsilon^{\rm syn}_{\rm k}\propto \gamma^2_{\rm k}$ for k=m,c)  and the electron Lorentz factors (eq. \ref{elec_Lorentz_c_h}),  the synchrotron spectral breaks and the maximum flux is
{\small
\bary\label{energies_break_syn_c_h}
\epsilon^{\rm syn}_{\rm m}&\simeq& 0.5\,{\rm GHz}\,\, \left(\frac{1+z}{1.022}\right)^{\frac{4-\alpha_s}{\alpha_s+8}}\,g(p)^{2} \varepsilon^2_{e,-1}\,\varepsilon_{B,-3}^{\frac12}\,n_{-1}^{\frac{\alpha_s}{2(\alpha_s+8)}}\,\tilde{E}^{\frac{4}{\alpha_s+8}}_{50}\, t_{\rm 1d}^{-\frac{12}{\alpha_s+8}} \hspace{1cm}\cr
\epsilon^{\rm syn}_{\rm c}&\simeq&  5.9\times 10^{-3}\,{\rm eV}  \left(\frac{1+z}{1.022}\right)^{\frac{\alpha_s-4}{\alpha_s+8}} (1+Y)^{-2}\, \varepsilon_{B,-3}^{-\frac32}\,n_{-1}^{-\frac{3\alpha_s+16}{2(\alpha_s+8)}}\,\tilde{E}^{-\frac{4}{\alpha_s+8}}_{50} \,\,t_{\rm 1d}^{-\frac{2(\alpha_s+2)}{\alpha_s+8}}\,, \cr
F^{\rm syn}_{\rm max} &\simeq& 1.9\times 10^{-1}\,{\rm mJy}\,\, \left(\frac{1+z}{1.022}\right)^{-\frac{4(\alpha_s+2)}{\alpha_s+8}}\,\varepsilon_{B,-3}^{\frac12}\,n_{-1}^{\frac{3\alpha_s+8}{2(\alpha_s+8)}}\, D^{-2}_{26.5}  \,\tilde{E}^{\frac{8}{\alpha_s+8}}_{50}\,t_{\rm 1d}^{\frac{3\alpha_s}{\alpha_s+8}}\,.
\eary
}

\paragraph{SSC emission.} From the electrons Lorentz factors (eqs. \ref{elec_Lorentz_c_h}) and synchrotron spectral breaks (eqs. \ref{energies_break_syn_c_h}), the SSC spectral break and the maximum flux is   
{\small
\bary\label{energies_break_ssc_c_h}
\epsilon^{\rm ssc}_{\rm m}&\simeq& 2.5\times 10^{-2}\,{\rm eV}\,\, \left(\frac{1+z}{1.022}\right)^{\frac{10-\alpha_s}{\alpha_s+8}}\, g(p)^4\,\varepsilon^4_{e,-1}\,\varepsilon_{B,-3}^{\frac12}\,n_{-1}^{\frac{\alpha_s-4}{2(\alpha_s+8)}}\,\tilde{E}^{\frac{6}{\alpha_s+8}}_{50}\, t_{\rm 1d}^{-\frac{18}{\alpha_s+8}} \hspace{1cm}\cr
\varepsilon^{\rm ssc}_{\rm c}&\simeq&  2.4\,{\rm GeV}  \left(\frac{1+z}{1.022}\right)^{\frac{3(\alpha_s-2)}{\alpha_s+8}} (1+Y)^{-4}\, \varepsilon_{B,-3}^{-\frac72}\,n_{-1}^{-\frac{7\alpha_s+36}{2(\alpha_s+8)}}\,\tilde{E}^{-\frac{10}{\alpha_s+8}}_{50}\,\,t_{\rm 1d}^{-\frac{2(2\alpha_s+1)}{\alpha_s+8}}\,, \cr
F^{\rm ssc}_{\rm max} &\simeq& 6.8\times 10^{-9}\,{\rm mJy}\,\, \left(\frac{1+z}{1.022}\right)^{-\frac{5(\alpha_s+2)}{\alpha_s+8}}\,g(p)^{-1}\varepsilon_{B,-3}^{\frac12}\,n_{-1}^{\frac{5(\alpha_s+4)}{2(\alpha_s+8)}}\, D^{-2}_{26.5}  \,\tilde{E}^{\frac{10}{\alpha_s+8}}_{50}\,t_{\rm 1d}^{\frac{2(\alpha_s+2)}{\alpha_s+8}}\,.
\eary
}
The break energy due to KN effect is 
\be
\varepsilon^{\rm ssc}_{\rm KN}\simeq 1.1 \,{\rm TeV}  \left(\frac{1+z}{1.022}\right)^{\frac{2\alpha_s-6}{\alpha_s+8}} (1+Y)^{-1}\, \varepsilon_{B,-3}^{-1}\,n_{-1}^{-\frac{\alpha_s+6}{\alpha_s+8}}\,\tilde{E}^{-\frac{2}{\alpha_s+8}}_{50}\,\,t_{\rm 1d}^{-\frac{\alpha_s+2}{\alpha_s+8}}\,.
\ee

\subsection{Lateral expansion}
Using the bulk Lorentz factor (eq. \ref{Gamma_c_l}),  we derive and show the observable quantities when the quasi-spherical outflow lies in the lateral expansion phase.

\paragraph{Synchrotron radiation.}   In this case, the minimum and cooling Lorentz factors are given by
{\small
\begin{eqnarray}\label{elec_Lorentz_c_h_l}
\gamma_{\rm m}&\simeq&  63.5\, \left(\frac{1+z}{1.022}\right)^{\frac{3}{\alpha_s+6}}\,g(p)\,\varepsilon_{e,-1}\,\,n_{-1}^{-\frac{1}{\alpha_s+6}}\,\beta^{-\frac{\alpha_s}{\alpha_s+6}}\tilde{E}^{\frac{1}{\alpha_s+6}}_{50} t_{\rm 30d}^{-\frac{3}{\alpha_s+6}}\cr
\gamma_{\rm c}&\simeq&  1.1\times 10^{5}\,\, \left(\frac{1+z}{1.022}\right)^{\frac{\alpha_s-3}{\alpha_s+6}}\,(1+Y)^{-1} \varepsilon^{-1}_{B,-1}\,\,n_{-1}^{-\frac{\alpha_s+3}{\alpha_s+6}}\,\beta^{\frac{3\alpha_s}{\alpha_s+6}}\tilde{E}^{-\frac{3}{\alpha_s+6}}_{50} t_{\rm 30d}^{\frac{3-\alpha_s}{\alpha_s+6}}\,.
\end{eqnarray}
}

Given the synchrotron radiation ($\epsilon^{\rm syn}_{\rm k}\propto \gamma^2_{\rm k}$ for k=m,c)  and eq. (\ref{elec_Lorentz_c_h_l}), the spectral break and the maximum flux of synchrotron radiation  is
{\small
\bary\label{energies_break_syn_c_h_l}
\epsilon^{\rm syn}_{\rm m}&\simeq& 0.3\times 10^{-2}\,{\rm GHz}\,\, \left(\frac{1+z}{1.022}\right)^{\frac{6-\alpha_s}{\alpha_s+6}} g(p)^2 \varepsilon^2_{e,-1}\,\varepsilon_{B,-3}^{\frac12}\,n_{-1}^{\frac{\alpha_s-2}{2(\alpha_s+6)}}\,\beta^{-\frac{4\alpha_s}{\alpha_s+6}}\tilde{E}^{\frac{4}{\alpha_s+6}}_{50}\, t_{\rm 30d}^{-\frac{12}{\alpha_s+6}} \hspace{1cm}\cr
\epsilon^{\rm syn}_{\rm c}&\simeq&  5.5 \times 10^{-3} \,{\rm keV}  \left(\frac{1+z}{1.022}\right)^{\frac{\alpha_s-6}{\alpha_s+6}} (1+Y)^{-2}\, \varepsilon_{B,-3}^{-\frac32}\,n_{-1}^{-\frac{3\alpha_s+10}{2(\alpha_s+6)}}\,\beta^{\frac{4\alpha_s}{\alpha_s+6}}\tilde{E}^{-\frac{4}{\alpha_s+6}}_{50}\,t_{\rm 30d}^{-\frac{2\alpha_s}{\alpha_s+6}}\,,\cr 
F^{\rm syn}_{\rm max} &\simeq& 4.3\,{\rm mJy}\,\, \left(\frac{1+z}{1.022}\right)^{-\frac{4\alpha_s}{\alpha_s+6}}\,\varepsilon_{B,-3}^{\frac12}\,n_{-1}^{\frac{3\alpha_s+2}{2(\alpha_s+6)}}\,\beta^{-\frac{8\alpha_s}{\alpha_s+6}} D^{-2}_{26.5}  \,\tilde{E}^{\frac{8}{\alpha_s+6}}_{50}\,t_{\rm 30d}^{\frac{3(\alpha_s-2)}{\alpha_s+6}}\,.
\eary
}

\paragraph{SSC emission.} From the electrons Lorentz factors (eqs. \ref{elec_Lorentz_c_h_l}) and synchrotron spectral breaks (eqs. \ref{energies_break_syn_c_h_l}), the SSC spectral break and the maximum flux is   
{\small
\bary\label{energies_break}
\epsilon^{\rm ssc}_{\rm m}&\simeq& 4.9\times 10^{-4}\,{\rm eV}\,\, \left(\frac{1+z}{1.022}\right)^{\frac{12-\alpha_s}{\alpha_s+6}}g(p)^4\varepsilon^4_{e,-1}\,\varepsilon_{B,-3}^{\frac12}\,n_{-1}^{\frac{\alpha_s-6}{2(\alpha_s+6)}}\,\beta^{-\frac{6\alpha_s}{\alpha_s+6}}\tilde{E}^{\frac{6}{\alpha_s+6}}_{50}\, t_{\rm 30d}^{-\frac{18}{\alpha_s+6}} \hspace{1cm}\cr
\epsilon^{\rm ssc}_{\rm c}&\simeq&  62.9\,{\rm GeV}  \left(\frac{1+z}{1.022}\right)^{\frac{3(\alpha_s-4)}{\alpha_s+6}} (1+Y)^{-4}\, \varepsilon_{B,-3}^{-\frac72}\,n_{-1}^{-\frac{7\alpha_s+22}{2(\alpha_s+6)}}\,\beta^{\frac{10\alpha_s}{\alpha_s+6}}\tilde{E}^{-\frac{10}{\alpha_s+6}}_{50}\,\,t_{\rm 30d}^{-\frac{2(2\alpha_s-3)}{\alpha_s+6}}\,, \cr
F^{\rm ssc}_{\rm max} &\simeq& 1.5\times 10^{-6}\,{\rm mJy}\,\, \left(\frac{1+z}{1.022}\right)^{-\frac{5\alpha_s}{\alpha_s+6}}\,g(p)^{-1}\varepsilon_{B,-3}^{\frac12}\,n_{-1}^{\frac{5(\alpha_s+2)}{2(\alpha_s+6)}}\, D^{-2}_{26.5}  \,\beta^{-\frac{10\alpha_s}{\alpha_s+6}}\tilde{E}^{\frac{10}{\alpha_s+6}}_{50}\,t_{\rm 30d}^{\frac{2(2\alpha_s-3)}{\alpha_s+6}}\,.
\eary
}

\subsection{Wind-like medium }
Using the bulk Lorentz factor (eq. \ref{Gamma_c_w}),  we derive and show the observable quantities when the quasi-spherical outflow is decelerated in a wind-like medium.
\paragraph{Synchrotron radiation.}   In this case, the minimum and cooling Lorentz factors are given by
{\small
\bary\label{ele_Lorentz_w}
\gamma_m&=& 5.1\times 10^2\, \left(\frac{1+z}{1.022}\right)^{\frac{1}{\alpha_s+4}} \,\xi^{-\frac{2}{\alpha_s+4}} g(p)  \varepsilon_{e,-1}\,A_{\star,-4}^{-\frac{1}{\alpha_s+4}}\,\tilde{E}^{\frac{1}{\alpha_s+4}}_{50}\, t_{\rm 10s}^{-\frac{1}{\alpha_s+4}}   \cr
\gamma_c&=&  3.7\times 10^4\, \left(\frac{1+z}{1.022}\right)^{-\frac{\alpha_s+3}{\alpha_s+4}}\,  (1+Y)^{-1}\, \,\xi^{\frac{2(2\alpha_s+7)}{\alpha_s+4}}   \varepsilon^{-1}_{B,-3}\,A_{\star,-4}^{-\frac{\alpha_s+5}{\alpha_s+4}}\,\tilde{E}^{\frac{1}{\alpha_s+4}}_{50}\, \,t_{\rm 10s}^{\frac{\alpha_s+3}{\alpha_s+4}}\,.
\eary
}

Given the synchrotron radiation ($\epsilon^{\rm syn}_{\rm k}\propto \gamma^2_{\rm k}$ for k=m,c)  and the electron Lorentz factors (eq. \ref{ele_Lorentz_w}),  the synchrotron spectral breaks and the maximum flux is
{\small
\bary\label{energies_break_syn_w}
\epsilon^{\rm syn}_{\rm m}&\simeq& 0.2\,{\rm eV}\,\, \left(\frac{1+z}{1.022}\right)^{\frac{2}{\alpha_s+4}}\, g(p)^{2}\,\xi^{-\frac{2(\alpha_s+6)}{\alpha_s+4}}\, \varepsilon^2_{e,-1}\,\varepsilon_{B,-3}^{\frac12}\,A_{\star,-4}^{\frac{\alpha_s}{2(\alpha_s+4)}}\,\tilde{E}^{\frac{2}{\alpha_s+4}}_{50}\, t_{\rm 10s}^{-\frac{(\alpha_s+6)}{\alpha_s+4}} \hspace{1cm}\cr
\epsilon^{\rm syn}_{\rm c}&\simeq&  3.7\,{\rm keV}  \left(\frac{1+z}{1.022}\right)^{-\frac{2(\alpha_s+3)}{\alpha_s+4}} \,\xi^{\frac{2(3\alpha_s+10)}{\alpha_s+4}}\, (1+Y)^{-2}\, \varepsilon_{B,-3}^{-\frac32}\,A_{\star,-4}^{-\frac{16+3\alpha_s}{2(\alpha_s+4)}}\,\tilde{E}^{\frac{2}{\alpha_s+4}}_{50}\,t_{\rm 10s}^{\frac{\alpha_s+2}{\alpha_s+4}}\,, \cr
F^{\rm syn}_{\rm max} &\simeq& 0.2\,{\rm mJy}\,\, \left(\frac{1+z}{1.022}\right)^{-\frac{\alpha_s+2}{\alpha_s+4}} \,\xi^{-\frac{4}{\alpha_s+4}}   \,\varepsilon_{B,-3}^{\frac12}\,A_{\star,-4}^{\frac{3\alpha_s+8}{2(\alpha_s+4)}}\, D^{-2}_{26.5}  \,\tilde{E}^{\frac{2}{\alpha_s+4}}_{50}\,t_{\rm 10s}^{-\frac{2}{\alpha_s+4}}\,.
\eary
}

\paragraph{SSC emission.} From the electron Lorentz factors (eqs. \ref{ele_Lorentz_w}) and synchrotron spectral breaks (eqs. \ref{energies_break_syn_w}), the SSC spectral break and the maximum flux is  
{\small
\bary\label{energies_break_ssc_w}
\epsilon^{\rm ssc}_{\rm m}&\simeq& 42.7\,{\rm keV}\,   \left(\frac{1+z}{1.022}\right)^{\frac{4}{\alpha_s+4}} \,g(p)^4 \xi^{-\frac{2(\alpha_s+8)}{\alpha_s+4}}\, \varepsilon^4_{e,-1}\,\varepsilon_{B,-3}^{\frac12}\,  A_{\star,-4}^{\frac{\alpha_s-4}{2(\alpha_s+4)}}\,\tilde{E}^{\frac{4}{\alpha_s+4}}_{50}\, t_{\rm 10s}^{-\frac{(\alpha_s+8)}{\alpha_s+4}} \hspace{1cm}\cr
\epsilon^{\rm ssc}_{\rm c}&\simeq&  5.4\,{\rm TeV}  \left(\frac{1+z}{1.022}\right)^{-\frac{4(\alpha_s+3)}{\alpha_s+4}}\,\xi^{\frac{2(7\alpha_s+24)}{\alpha_s+4}}\, (1+Y)^{-4}\, \varepsilon_{B,-3}^{-\frac72}\,A_{\star,-4}^{-\frac{7\alpha_s+36}{2(\alpha_s+4)}}\,\tilde{E}^{\frac{4}{\alpha_s+4}}_{50}\,\,t_{\rm 10s}^{\frac{3\alpha_s+8}{\alpha_s+4}}\,, \cr
F^{\rm ssc}_{\rm max} &\simeq& 2.2\times 10^{-7}\,{\rm mJy}\,\,g(p)^{-1}\xi^{-2}\varepsilon_{B,-3}^{\frac12}\,A_{\star,-4}^{\frac{5}{2}}\, D^{-2}_{26.5}\,t_{\rm 10s}^{-1}\,.
\eary
}
The break energy due to KN effect is 
\be
\epsilon^{\rm ssc}_{\rm KN}\simeq 313.9 \,{\rm GeV}  \left(\frac{1+z}{1.022}\right)^{-\frac{2\alpha_s+6}{\alpha_s+4}}\,\xi^{\frac{2(2\alpha_s+6)}{\alpha_s+4}}\, (1+Y)^{-1}\, \varepsilon_{B,-3}^{-1}\,A_{\star,-4}^{-\frac{\alpha_s+6}{\alpha_s+4}}\,\tilde{E}^{\frac{2}{\alpha_s+4}}_{50}\,\,t_{\rm 10s}^{\frac{\alpha_s+2}{\alpha_s+4}}\,.
\ee

\newpage
\section{B. Off-axis jet}
\subsection{Homogeneous medium}
Using the bulk Lorentz factor (eq. \ref{Gamma_o}),  we derive and show the observable quantities when the off-axis jet is decelerated in a homogeneous medium.

\paragraph{Synchrotron radiation.} In tis case, the minimum and cooling electron Lorentz factors are  given by
{\small
\begin{eqnarray}\label{elec_Lorentz_o_h}
\gamma_{\rm m}&=& 6.9\times 10^3\, \varepsilon_{e,-1} \left(\frac{1+z}{1.022}\right)^{\frac{3}{2}}\,g(p) \,n_{-4}^{-\frac{1}{2}} \,\tilde{E}_{50}^{\frac{1}{2}}\,\Delta\theta_{15^\circ}^{3}\,\theta_{j,5^\circ}^{-1}\,t_{\rm 1d}^{-\frac{3}{2}}\,,\cr
\gamma_{\rm c}&=& 8.7\times 10^3\,\, \left(\frac{1+z}{1.022}\right)^{-\frac{1}{2}}\,(1+Y)^{-1} \varepsilon^{-1}_{B,-4}\,n_{-1}^{-\frac{1}{2} }\,E^{-\frac{1}{2}}_{50}\,\Delta\theta_{15^\circ}^{-1}\,\theta_{j,5^\circ}\,t_{\rm 1d}^{\frac{1}{2}}\,.
\end{eqnarray}
}

Given the synchrotron radiation ($\epsilon^{\rm syn}_{\rm k}\propto \gamma^2_{\rm k}$ for k=m,c)  and the electron Lorentz factors (eq. \ref{elec_Lorentz_o_h}),  the synchrotron spectral breaks and the maximum flux is
{\small
\bary\label{energies_break_syn_off_h}
\epsilon^{\rm syn}_{\rm m}&\simeq& 4.7\times 10^{3}\,{\rm GHz}\,\, \left(\frac{1+z}{1.022}\right)^2\,g(p)^{2} \varepsilon^2_{e,-1}\,\varepsilon_{B,-3}^{\frac12}\,n_{-1}^{-\frac{1}{2}}\,E_{50}\,\Delta\theta_{15^\circ}^{4}\,\theta_{j,5^\circ}^{-2}\,  t_{\rm 1d}^{-3} \hspace{1cm}\cr
\epsilon^{\rm syn}_{\rm c}&\simeq&  8.1\times 10^{-3}\,{\rm keV}  \left(\frac{1+z}{1.022}\right)^{-2} (1+Y)^{-2}\, \epsilon_{B,-3}^{-\frac32}\,n_{-1}^{-\frac{1}{2}}\,E^{-1}_{50}\Delta\theta_{15^\circ}^{-4}\theta_{j,5^\circ}^{2}\,\,t_{\rm 1d}\,,\cr 
F^{\rm syn}_{\rm max} &\simeq& 3.5\times 10^{-2}\,{\rm mJy}\,\, \left(\frac{1+z}{1.022}\right)^{-4}\,\varepsilon_{B,-3}^{\frac12}\,n_{-1}^{\frac{5}{2}}\, D^{-2}_{26.5}  \,E^{-1}_{50}\,\Delta\theta_{15^\circ}^{-18}\,\theta_{j,5^\circ}^{2}\, t_{\rm 1d}^{6}\,.
\eary
}

\paragraph{SSC emission} From the electron Lorentz factors (eqs. \ref{elec_Lorentz_c_h_l}) and synchrotron spectral breaks (eqs. \ref{energies_break_syn_off_h}), the SSC spectral break and the maximum flux is  
{\small
\bary\label{energies_break_ssc_off_h}
\epsilon^{\rm ssc}_{\rm m}&\simeq& 0.9\,{\rm MeV}\,\left(\frac{1+z}{1.022}\right)^{5}\, g(p)^{4}\varepsilon^4_{e,-1}\,\varepsilon_{B,-3}^{\frac12}\,n_{-4}^{-\frac{3}{2}}\,E^2_{50}\,\Delta\theta_{15^\circ}^{10}\,\theta_{j,5^\circ}^{-4}\,  t_{100\,{\rm d}}^{-6} \hspace{1cm}\cr
\epsilon^{\rm ssc}_{\rm c}&\simeq&  0.6\,{\rm MeV}  \left(\frac{1+z}{1.022}\right)^{-3} (1+Y)^{-4}\, \varepsilon_{B,-3}^{-\frac72}\,n_{-4}^{-\frac{3}{2}}\,E^{-2}_{50}\Delta\theta_{15^\circ}^{-6}\theta_{j,5^\circ}^{4}\,t^2_{100\,{\rm d}}\,,\cr 
F^{\rm ssc}_{\rm max} &\simeq& 2.7\times 10^{-9}\,{\rm mJy}\,\, \left(\frac{1+z}{1.022}\right)^{-5}\,g(p)^{-1}\varepsilon_{B,-3}^{\frac12}\,n_{-4}^{\frac{7}{2}}\, D^{-2}_{26.5}  \,E^{-1}_{50}\,\Delta\theta_{15^\circ}^{-20}\,\theta_{j,5^\circ}^{2}\, t_{100\,{\rm d}}^{7}\,.
\eary
}
The break energy due to KN effect is 
\be
\epsilon^{\rm ssc}_{\rm KN}\simeq 1.4 \,{\rm TeV}  (1+Y)^{-1}\, \varepsilon_{B,-3}^{-1}\,n_{-4}^{-1}\,\Delta\theta_{15^\circ}^{2}\,t^{-1}_{100\,{\rm d}}\,.
\ee

\subsection{Lateral expansion}
Using the bulk Lorentz factor (eq. \ref{Gamma_j_l}),  we derive and show the observable quantities when the off-axis jet lies in the lateral expansion phase.

\paragraph{Synchrotron radiation.}  In this case, the minimum and cooling Lorentz factors are given by
{\small
\begin{eqnarray}\label{elec_Lorentz_o_h_l}
\gamma_{\rm m}&\simeq&  50.2\, \left(\frac{1+z}{1.022}\right)^{\frac{1}{2}}\,g(p)\,\varepsilon_{e,-1}\,\,n_{-1}^{-\frac{1}{6}}\,\tilde{E}^{\frac{1}{6}}_{50} t_{\rm 1d}^{-\frac{1}{2}}\cr
\gamma_{\rm c}&\simeq&  1.3\times 10^{5}\,{\rm GHz}\,\, \left(\frac{1+z}{1.022}\right)^{-\frac{1}{2}}\,(1+Y)^{-1} \varepsilon^{-1}_{B,-1}\,\,n_{-1}^{-\frac{1}{2}}\,\tilde{E}^{-\frac{1}{2}}_{50} t_{\rm 1d}^{\frac{1}{2}}\,.
\end{eqnarray}
}

Given the synchrotron radiation ($\epsilon^{\rm syn}_{\rm k}\propto \gamma^2_{\rm k}$ for k=m,c)  and eq. (\ref{elec_Lorentz_o_h_l}), the spectral break and the maximum flux of synchrotron radiation is
{\small
\bary\label{energies_break_syn_off_h_l}
\epsilon^{\rm syn}_{\rm m}&\simeq& 1.2\times 10^{-2}\,{\rm GHz}\, \left(\frac{1+z}{1.022}\right)\,g(p)^{2}\,\varepsilon^2_{e,-1}\,\varepsilon_{B,-3}^{\frac12}\,n_{-1}^{-\frac{1}{6}}\,  E^{\frac{2}{3}}_{50}\, t_{\rm 1d}^{-2}\,\,\,\, \cr
\epsilon^{\rm syn}_{\rm c}&\simeq&  1.4\,{\rm eV} \, \left(\frac{1+z}{1.022}\right)^{-1}    (1+Y)^{-2}\, \varepsilon_{B,-3}^{-\frac32}\,n_{-1}^{-\frac{5}{6}}  \,E^{-\frac{2}{3}}_{50}\cr
F^{\rm syn}_{\rm max} &\simeq& 24.2\,{\rm mJy}\, \left(\frac{1+z}{1.022}\right)^3\,\varepsilon_{B,-3}^{\frac12}\,n_{-1}^{\frac{1}{6}}\,D^{-2}_{26.5}  \,E^{\frac{4}{3}}_{50}\,t_{\rm 1d}^{-1}.\cr
&&\hspace{5cm}
\eary
}

\paragraph{SSC radiation.} From the electron Lorentz factors (eqs. \ref{elec_Lorentz_o_h_l}) and synchrotron spectral breaks (eqs. \ref{energies_break_syn_off_h_l}), the SSC spectral break and the maximum flux is  
{\small
\bary\label{energies_break_ssc_off_h_l}
\epsilon^{\rm ssc}_{\rm m}&\simeq& 1.4\times10^{-4}\,{\rm eV}\,\, \left(\frac{1+z}{1.022}\right)^2\,g(p)^4 \epsilon^4_{e,-1}\,\varepsilon_{B,-3}^{\frac12}\,n_{-1}^{-\frac{1}{2}}\,E_{50}\,t_{\rm 1d}^{-3} \hspace{1cm}\cr
\epsilon^{\rm ssc}_{\rm c}&\simeq&  23.4\,{\rm GeV}  \left(\frac{1+z}{1.022}\right)^{-2} (1+Y)^{-4}\, \varepsilon_{B,-3}^{-\frac72}\,n_{-1}^{-\frac{11}{6}}\,E^{-\frac53}_{50}\,t_{\rm 1d}\,,\cr 
F^{\rm ssc}_{\rm max} &\simeq& 1.7\times 10^{-5}\,{\rm mJy}\,\, \left(\frac{1+z}{1.022}\right)^3\,g(p)^{-1}\varepsilon_{B,-3}^{\frac12}\,n_{-1}^{\frac56}\, D^{-2}_{26.5}  \,E^{\frac53}_{50}\, t_{\rm 1d}^{-1}\,.
\eary
}

\subsection{Wind-like medium}
Using the bulk Lorentz factor (eq. \ref{Gamma_o_w}),  we derive and show the observable quantities when the off-axis jet is decelerated in a wind-like medium.

\paragraph{Synchrotron radiation.}   In this case, the minimum and cooling Lorentz factors are given by
{\small
\bary\label{ele_Lorentz_w_o}
\gamma_m&=& 8.7\times 10^4\, \left(\frac{1+z}{1.022}\right)^{\frac{1}{2}} \,\xi^{-1} g(p)  \varepsilon_{e,-1}\,A_{\star,-4}^{-\frac{1}{2}}\,\theta_{j,5^\circ}^{-1}\Delta\theta_{15^\circ}\tilde{E}^{\frac{1}{2}}_{50}\, t_{\rm 10s}^{-\frac{1}{2}}   \cr
\gamma_c&=&  1.5\times 10^{-3}\, \left(\frac{1+z}{1.022}\right)^{-\frac{3}{2}}\,  (1+Y)^{-1}\, \,\xi^{3}   \varepsilon^{-1}_{B,-3}\,A_{\star,-4}^{-\frac{1}{2}}\,\theta_{j,5^\circ}\Delta\theta_{15^\circ}^{-3}\tilde{E}^{-\frac{1}{2}}_{50}\, \,t_{\rm 10s}^{\frac{3}{2}}\,.
\eary
}

Given the synchrotron radiation ($\epsilon^{\rm syn}_{\rm k}\propto \gamma^2_{\rm k}$ for k=m,c)  and the electron Lorentz factors (eq. \ref{ele_Lorentz_w_o}),  the synchrotron spectral breaks and the maximum flux is
{\small
\bary\label{energies_break_syn_off_w}
\epsilon^{\rm syn}_{\rm m}&\simeq& 95.5\,{\rm keV}\,\, \left(\frac{1+z}{1.022}\right)\,g(p)^{2}\xi^{-4}\, \varepsilon^2_{e,-1}\,\varepsilon_{B,-3}^{\frac12}\,A_{\star,-4}^{-\frac{1}{2}}\,E_{50}\,\Delta\theta_{15^\circ}^2\,\theta_{j,5^\circ}^{-2}\,  t_{\rm 10s}^{-2} \hspace{1cm}\cr
\epsilon^{\rm syn}_{\rm c}&\simeq&  8.4\times 10^{-13}\,{\rm eV}  \left(\frac{1+z}{1.022}\right)^{-3}\,\xi^4\, (1+Y)^{-2}\, \varepsilon_{B,-3}^{-\frac32}\,A_{\star,-4}^{-\frac{1}{2}}\,E^{-1}_{50}\Delta\theta_{15^\circ}^{-6}\theta_{j,5^\circ}^{2}\,\,t^2_{\rm 10s}\,,\cr 
F^{\rm syn}_{\rm max} &\simeq& 0.2\,{\rm mJy}\,\, \left(\frac{1+z}{1.022}\right)\, \xi^2\,\varepsilon_{B,-3}^{\frac12}\,A_{\star,-4}^{\frac{5}{2}}\, D^{-2}_{26.5}  \,E^{-1}_{50}\,\Delta\theta_{15^\circ}^{-8}\,\theta_{j,5^\circ}^{2}\, t_{\rm 10s}\,.
\eary
}

\paragraph{SSC emission} From the electron Lorentz factors (eqs. \ref{ele_Lorentz_w_o}) and synchrotron spectral breaks (eqs. \ref{energies_break_syn_off_w}), the SSC spectral break and the maximum flux is  
{\small
\bary\label{energies_break_ssc_off_w}
\epsilon^{\rm ssc}_{\rm m}&\simeq& 3.1\,{\rm GeV}\,\, \left(\frac{1+z}{1.022}\right)^2\,g(p)^{4}\xi^{-6}\, \varepsilon^4_{e,-1}\,\varepsilon_{B,-3}^{\frac12}\,A_{\star,-4}^{-\frac{3}{2}}\,E^2_{50}\,\Delta\theta_{15^\circ}^{4}\,\theta_{j,5^\circ}^{-4}\,  t_{\rm 10s}^{-3} \hspace{1cm}\cr
\epsilon^{\rm ssc}_{\rm c}&\simeq&  1.8\times 10^{-18}\,{\rm eV}  \left(\frac{1+z}{1.022}\right)^{-6}\,\xi^{10}\, (1+Y)^{-4}\, \varepsilon_{B,-3}^{-\frac72}\,A_{\star,-4}^{-\frac{3}{2}}\,E^{-2}_{50}\Delta\theta_{15^\circ}^{-12}\theta_{j,5^\circ}^{4}\,\,t^5_{\rm 10s}\,,\cr 
F^{\rm ssc}_{\rm max} &\simeq& 3.1\,{\rm mJy}\,\, \left(\frac{1+z}{1.022}\right)^2\,g(p)^{-1}\varepsilon_{B,-3}^{\frac12}\,A_{\star,-4}^{\frac{7}{2}}\, D^{-2}_{26.5}  \,E^{-1}_{50}\,\Delta\theta_{15^\circ}^{-6}\,\theta_{j,5^\circ}^{2}\,\,.
\eary
}
The break energy due to KN effect is 
\be
\epsilon^{\rm ssc}_{\rm KN}\simeq 2.4\times 10^{-3} \,{\rm GeV}  \left(\frac{1+z}{1.022}\right)^{-2}\,\xi^{2}\, (1+Y)^{-1}\, \varepsilon_{B,-3}^{-1}\,A_{\star,-4}^{-1}\,\Delta\theta_{15^\circ}^{-2}\,t_{\rm 10s}\,.
\ee

\end{document}